\documentclass[aps,pra, two column,10pt,superscriptaddress,longbibliography,nofootinbib]{revtex4-2}

\usepackage{etex}
\usepackage{physics}
\usepackage{amsmath,amssymb,amsthm,amstext,mathrsfs}
\usepackage[colorlinks=true,citecolor=blue,urlcolor=blue]{hyperref}
\usepackage{comment}
\usepackage{eufrak}
\usepackage[pdftex]{graphicx}
\usepackage{enumerate}
\usepackage{times,txfonts}
\usepackage{braket}
\usepackage{svg}
\usepackage{color}
\usepackage{natbib}
\usepackage{amsmath,blkarray}
\usepackage{mathtools}
\usepackage{latexsym}
\usepackage{tabularx, booktabs}
\usepackage{graphics,epstopdf}
\usepackage{graphicx, lipsum}
\usepackage{float}
\usepackage{amsfonts}
\usepackage{subcaption}
\usepackage{enumerate}
\usepackage{color,soul}

\newcommand{\be}{\begin{equation}}
	\newcommand{\ee}{\end{equation}}
\newcommand{\ba}{\begin{eqnarray}}
	\newcommand{\ea}{\end{eqnarray}}

\newtheorem{thm}{Theorem}
\newtheorem{Lemma}{Lemma}

\definecolor{ss}{RGB}{250,80,220}

\begin{document}

\title{Self-testing of multiple unsharpness parameters through sequential violations of non-contextual inequality}

\author{Rajdeep Paul}
\email{ph21resch11016@iith.ac.in}
\affiliation{Indian Institute of Technology Hyderabad, Kandi, Sangareddy, Telangana 502285, India}

\author{Souradeep Sasmal}
\email{souradeep.007@gmail.com}
\affiliation{Institute of Fundamental and Frontier Sciences, University of Electronic Science and Technology of China, Chengdu 611731, China}
\affiliation{Indian Institute of Technology Hyderabad, Kandi, Sangareddy, Telangana 502285, India}

\author{A. K. Pan }
\email{akp@phy.iith.ac.in}
\affiliation{Indian Institute of Technology Hyderabad, Kandi, Sangareddy, Telangana 502285, India}

\begin{abstract}
The self-testing protocols refer to novel device-independent certification schemes wherein the devices are uncharacterised, and the dimension of the system remains unspecified.  The optimal quantum violation of a Bell's inequality facilitates such self-testing.  In this work,  we put forth a protocol for self-testing of noisy quantum instruments, specifically, the unsharpness parameter of smeared projective measurements in any arbitrary dimension. Our protocol hinges on the sequential quantum violations of a bipartite Bell-type preparation non-contextual inequality, involving three measurement settings per party. First, we demonstrate that at most three sequential independent Bobs manifest simultaneous preparation contextuality with a single Alice through the violation of this inequality. Subsequently, we show that the sub-optimal sequential quantum violations of the non-contextual inequality form an optimal set, eventually enabling the self-testing of shared state, local measurements and unsharpness parameters of one party. Notably, we derive the optimal set of quantum violations without specifying the dimension of the quantum system, thereby circumventing the constraint that may arise due to Naimark's theorem. Furthermore, we extend our investigation to quantify the degree of incompatible measurements pertaining to the sequential observers, exploring how variations in the degree of incompatibility impact the values of unsharp parameters necessary for sequential quantum violation.  

\end{abstract}
\maketitle

%%%%%%%%%%%%%%%%%%%%%%%%%%%%%%%%%%%%%%%%%%%%%%%%%%%%%%%%%%%%%%%%%%%%%%%%%%%%%%%%%%%%%%%%%%%%%%%%%%%%%%%%%%%%%%%%%%%%%%%%%%%%%%%%%%%%%%%%%%%%%%%%%%%%%%%%%%%%%%%%%%%%%%%%%%%%%%%%%%%%%%%%%%%%%%%%%%%%%%%%%%%%%%%%%%%%%%%%%%%%%%%%%%%%%%%%%%%%%%%%%%%%%%%%%%%%%%%%%%%%%%%%%%%%%%%%%%%%%%%%%%%%%%%%%%%%%%%%%%%%%%%%%%%%%%%%%%%%%%%%%%%%%%%%%%%%%%%%%%

\section{Introduction}

The self-testing represents a novel approach that facilitates the strongest possible form of device-independent (DI) certification of the quantum systems solely from the observed input-output statistics \cite{Supic2020rev}. In such a protocol, the quantum devices are considered to be black boxes and the dimension of the quantum system are unknown.  The DI self-testing relies on the optimal quantum violation of a suitable Bell's inequality \cite{Bell1964}  enabling unique characterisation of state and measurements.  For instance, the optimal violation of the Clauser-Horn-Shimony-Halt (CHSH) inequality \cite{Clauser1969} self-tests the bipartite state to be maximally entangled, and local observables are anti-commuting \cite{Roy2023}. Besides a plethora of applications in information theoretic tasks, the DI self-testing provides foundational insights into understanding the geometric structure of the set of quantum correlations. It is worthwhile to note that the optimal quantum violation of a Bell inequality signifies that the quantum correlation in question is an extremal point of the set of all quantum correlations.

Since the inception of the self-testing protocol by Mayers and Yao \cite{mayers1998}, a flurry of protocols has been proposed, including parallel self-testing of multiple maximally entangled two-qubit states \cite{McKague2016, Wu2016}, self-testing of the pure non-maximally entangled two-qubit state \cite{Acin2012, Bamps2015, Coladangelo2017, Rai2021, Wooltron2022, Rai2022}. Recent developments have extended self-testing protocols to multipartite scenarios. Using specific linear and quadratic Bell inequalities, self-testing of $N$-partite GHZ state and anti-commuting observables for each party has been demonstrated \cite{Panwar2023}. Additionally, self-testing of tripartite W-state has been proposed utilising the SWAP-circuit method \cite{Wu2014}. Moreover, self-testing protocols have been extended for higher dimensional states, such as maximally entangled two-qudit states \cite{Sarkar2021}, as well as for multipartite graph states \cite{McKague2014} and optimal states for XOR games\cite{miller2013}. Self-testing of measurements and inputs has been reported in \cite{Gomez2016, supic2020input}. Quite a number of works on self-testing of the non-projective measurements in device-independent or semi-device-independent scenarios, where either dimension of the quantum system is used or in prepare measure scenario \cite{Tavakoli2020sc, Smania2020, Miklin2020, Gomez2016, Gomez2018, Pan2021}. Recently, device-independent certification of an unsharp instrument was also reported \cite{Roy2023}. In the experiment scenario, state and measurements have been self-tested in \cite{Mironowicz2019}.

Note that, the Bell's inequality is a test of a notion of classicality widely known as local realism. A distinct perspective on classicality emerged through the works of Kochen and Specker \cite{Kochen1967}, and later was generalized by Spekkens \cite{Spekkens2005}, who framed the notion of classicality in terms of non-contextuality.  In this regard, recent developments support \cite{Tavakoli2020jan, Roy2022}  that the non-contextuality may constitute a more fundamental notion of classicality than that of local realism. This assertion is underpinned by establishing the connection between steering with preparation noncontextuality via measurement incompatibility and hence concluding noncontextual realist models as a subset of local realist models \cite{Tavakoli2020jan, Roy2022, Plavala2022}. This implies that even when a quantum correlation has an underlying local realist model, it may manifest non-classicality in the form of contextuality. Apart from the immense foundational significance the nonlocality and contextuality poses, the latter, much like nonlocal correlations \cite{Brunner2014, Supic2020rev}, emerges as a pivotal resource with diverse applications in the realm of information processing such as communication games \cite{Spekkens2009, Hameedi2017, Pan2019, Pan2021}, state discrimination \cite{Schmid2018, Gomez2022, Flatt2022}, semi device-independent randomness certification \cite{Carceller2022}, sequential sharing of correlations \cite{Kumari2023}, self-testing protocols \cite{Saha2020, Abhyoudai2023}, and quantum computation \cite{Vega2017, Frembs2018, Bravyi2018}.

In this paper, we aim to self-test noisy quantum instruments, specifically the unsharp parameter of the smeared version of projective measurement, through the sequential quantum violations of a preparation non-contextual inequality. Note that, the unsharp measurement in a standard Bell experiment produces sub-optimal quantum violation and hence any kind of certification becomes challenging. The sub-optimal quantum violation in an experiment may arise from various factors, such as (i) non-ideal preparation of the state than that is intended, (ii) the implementation of local observables is inappropriate, or  (iii) the presence of noise in implementing of projective measurements, which can be modelled as unsharp measurements \cite{Busch1986, Busch1987}. Hence, to self-test an unsharp quantum instrument inevitably requires the simultaneous DI certification of the state, observables, and unsharpness parameter from the same observed input-output statistics. Note that, the effect of the unsharp measurements is reflected in the post-measurement states and the standard Bell test is incapable of certifying them. We invoke sequential Bell experiments which have the potential to self-test the unsharpness parameter along with the state and measurements as the sequential quantum violations depend on the post-measurement states. However, such a sequential quantum violations are sub-optimal and the challenge is to certify the state, measurements and unsharpness parameter from a set of sub-optimal quantum violations. Note that, the sequential sharing of various forms of quantum correlation by multiple sequential observers has been extensively explored \cite{Silva2015, Sasmal2018, Brown2020, Steffinlongo2022, Kumari2023}. Our work can then be viewed as a potential application of such studies.  
 
A Bell test involves two distant parties, Alice and Bob who share a bipartite entangled state $\rho_{1} \in \mathscr{H}^d\otimes \mathscr{H}^d$, where $d$ is the dimension of the local system. A sequential Bell experiment involves a single Alice (who always performs sharp projective measurements of her two observables)  and an arbitrary $k$ number of sequential Bobs (say, Bob$^{k}$) who performs unsharp measurements. The $k^{th}$ Bob may perform sharp measurement. After performing unsharp measurements on their respective subsystems, Bob$^{1}$ passes his residual subsystem to another sequential observer Bob$^{2}$ and so forth.  The process continues until Alice and Bob$^{k}$ get the violation of Bell's inequality. The choice of unsharp measurement is imperative as projective measurements maximally disturb the system, thereby destroying the entanglement and hence sequential Bobs has no chance to get the violation of Bell's inequality.

To this end, it is worth noting here that recently, semi-DI certification of an unsharp instrument has been demonstrated both theoretically \cite{Mohan2019, Mukherjee2021, Miklin2020, shi2019, Tavakoli2020sdic, Farkas2019, Foletto2020, Abhyoudai2023} and experimentally \cite{Anwer2020, Li2023} by assuming qubit system. Following this, quantification of the degree of incompatibility of two sequential pairs of quantum measurements has been demonstrated through the sequential quantum advantage \cite{Anwer2020}. The first DI certification of an unsharp instrument has been proposed \cite{Roy2023} based on CHSH inequality \cite{Clauser1969}.

In this work, we demonstrate how sub-optimal quantum violations form an optimal pair (or tuple) to self-test the unsharpness parameters of Bob$^1$. It was believed that the DI self-testing of the unsharp parameter is not possible due to Naimark's theorem which states that any non-projective measurement can be interpreted as the projective measurement in higher dimensional space. Since in the DI scenario, no dimension restriction is imposed, a stubborn individual always argues that the measurement is projective, but the experimentalist could not prepare the ideal observables required for the optimal quantum value. We overcome this by considering the dimension-independent optimisation of sequential Bell values simultaneously. We invoke an elegant sum-of-square (SOS) approach to evaluate the sub-optimal quantum value of the Bell functional in sequential scenario, without assuming the dimension of the system. We demonstrate that maximum of three independent sequential observers (Bob$_1$, Bob$_2$ and Bob$_3$) can demonstrate the sub-optimal quantum violation of the inequality with a single Alice. This enables us to robustly certify the unsharpness parameters of Bob$_1$ and Bob$_2$.

Note that incompatible measurements are necessary for demonstrating the preparation contextual quantum correlations \cite{Tavakoli2020jan, Roy2022}. By quantifying the degree of incompatibility pertaining to Bob's observables, we establish a quantitative relationship between the degree of incompatibility and the sequential quantum violations of Bell's inequality. We note that such a relationship was introduced in \cite{Anwer2020}, where the authors demonstrated how the degree of incompatibility for each pair of Bob's measurements leads to sequential sub-optimal violations of the CHSH inequality. However, they \cite{Anwer2020} considered the qubit system for their demonstration. We first show that such a relationship can be demonstrated without assuming the dimension of the quantum system. The sequential quantum violations of Bell's inequality certify unsharp parameters, enabling the certification of the degree of incompatibilities of each sequential Bob's observables. 
Further, we extend the analysis to trine observables based on the sequential quantum violations of our non-contextual inequality. By using the degree of incompatibility for three observables defined in \cite{Pal2011, Busch2014}, we quantitatively demonstrate how the degree of incompatibility for each sequential Bob can be certified from the sub-optimal violations of the non-contextual inequality. Finally, we wrap up with concluding remarks and a brief discussion of future possibilities in Sec.~\ref{SO}.

%%%%%%%%%%%%%%%%%%%%%%%%%%%%%%%%%%%%%%%%%%%%%%%%%%%%%%%%%%%%%%%%%%%%%%%%%%%%%%%%%%%%%%%%%%%%%%%%%%%%%%%%%%%%%%%%%%%%%%%%%%%%%%%%%%%%%%%%%%%%%%%%%%%%%%%%%%%%%%%%%%%%%%%%%%%%%%%%%%%%%%%%%%%%%%%%%%%%%%%%%%%%%%%%%%%%%%%%%%%%%%%%%%%%%%%%%%%%%%%%%%%%%%%%%%%%%%%%%%%%%%%%%%%%%%%%%%%%%%%%%%%%%%%%%%%%%%%%%%%%%%%%%%%%%%%%%%%%%%%%%%%%%%%%%%%%%%%%%%%%%%%%%%%%%%%%%%%%%%%%%%%%%%%%%%%%%%%%%%%%%%%%%%%%%%%%%%%%%%%%%%%%%%%%%%%%%%%%%%%%%%%%%%%%%%%%%%%

\section{A Bell functional and its preparation noncontextual bound}\label{BIC}
We provide a brief overview of the ontological model of operational quantum theory and the notion of classicality in terms of noncontextuality. We consider a Bell functional whose upper bound in a preparation noncontextual model is lower than the local bound.  

%%%%%%%%%%%%%%%%%%%%%%%%%%%%%%%%%%%%%%%%%%%%%%%%%%%%%%%%%%%%%%%%%%%%%%%%%%%%%%%%%%%%%%%%%%%%%%%%%%%%%%%%%%%%%%%%%%%%%%%%%%%%%%%%%%%%%%%%%%%%%%%%%%%%%%%%%%%%%%%%%%%%%%%%%%%%%%%%%%%%%%%%%%%%%%%%%%%%%%%%%%%%%%%%%%%%%%%%%%%%%%%%%%%%%%%%%%%%%%%%%%%%%%%%%%%%%%%%

\subsection{Ontological model and preparation noncontextuality}

 The primitives of an operational theory are a set of preparation procedures denoted as $\qty{\mathbb{P}}$ and a set of measurement procedures denoted as $\qty{\mathbb{M}}$. The probability of obtaining a specific outcome $k$ is given by $p(k|\mathcal{P}, \mathcal{M})$. In operational quantum theory, a specific preparation procedure $\mathcal{P}\in \mathbb{P}$ prepares the quantum state $\rho$ and the measurements $\mathcal{M}\in \mathbb{M}$ performed on the quantum state are in general characterised by a set of positive semi-definite operators, known as positive-operator-valued measures (POVMs), denoted as $\qty{\mathcal{E}_k}$, satisfying $\sum_k \mathcal{E}_k=\openone$. The quantum probability is obtained from the Born rule, $p(k|\mathcal{M},\rho)=\Tr[\rho \mathcal{E}_k]$. 

In an ontological model of quantum theory the preparation of $\rho$ through a specific procedure yields that an ontic state $\lambda\in\Lambda$ with a probability distribution $\mu(\lambda|\mathcal{P})$ satisfying $\sum_{\Lambda} \mu(\lambda|\mathcal{P}) =1$, where $\Lambda$ denotes the ontic state space. The probability of obtaining an outcome $k$ is given by a response function $\mathscr{E}(k|\lambda, \mathcal{E}_k)$ with $\sum_k \mathscr{E}(k|\lambda, \mathcal{E}_k)=1 \ \forall \lambda$. Any ontological model that is consistent with quantum theory must reproduce the Born rule, i.e,
\begin{equation}
    \sum_{\Lambda} \mu(\lambda|\rho, \mathcal{P}) \ \mathscr{E}(k|\lambda, \mathcal{M}) = \Tr[\rho \mathcal{E}_k] \nonumber
\end{equation}

An ontological model of the operational theory is said to be preparation non-contextual \cite{Spekkens2005} if two preparation procedures $\mathcal{P}_1$ and $\mathcal{P}_2$ yielding the \textit{same} quantum state $\rho$ that cannot be operationally distinguished by any measurement, i.e., 
\begin{equation}
p(k|\mathcal{P}_1,\mathcal{M})=p(k|\mathcal{P}_2,\mathcal{M}) \implies \mu(\lambda|\rho,\mathcal{P}_1)=\mu(\lambda|\rho,\mathcal{P}_2) \forall \lambda \nonumber 
\end{equation}
This means that in a preparation noncontextual ontic model, ontic state distributions are equivalent, irrespective of the preparation procedures.

%%%%%%%%%%%%%%%%%%%%%%%%%%%%%%%%%%%%%%%%%%%%%%%%%%%%%%%%%%%%%%%%%%%%%%%%%%%%%%%%%%%%%%%%%%%%%%%%%%%%%%%%%%%%%%%%%%%%%%%%%%%%%%%%%%%%%%%%%%%%%%%%%%%%%%%%%%%%%%%%%%%%%%%%%%%%%%%%%%%%%%%%%%%%%%%%%%%%%%%%%%%%%%%%%%%%%%%%%%%%%%%%%%%%%%%%%%%%%%%%%%%%%%%%%%%%%%%%

\subsection{Bipartite Bell functional for three inputs per party}

We consider a Bell experiment featuring two spatially separated parties, Alice and Bob. Alice (Bob) randomly performs one of three dichotomic local measurements $A_x\in \{A_1, A_2, A_{3}\}$ $(B_y\in \{B_1, B_2, B_3\})$. The respective measurement outcomes are denoted by $a,b \in \{+1, -1\}$. Given the above Bell scenario, consider the following Bell functional
\begin{eqnarray} \label{nbell}
\mathcal{I}&=&(A_1+A_2-A_3)\otimes B_1+(A_1-A_2+A_3)\otimes B_2 \nonumber \\
&& +(-A_1+A_2+A_3)\otimes B_3
\end{eqnarray}

The quantum value of this Bell functional is given by $\mathscr{I}=\Tr[\mathcal{I} \rho]$. In an ontological model, the local bound of the Bell functional is $\mathcal{I}_{l}\leq 5$ \cite{Gisin1999}. Note that while by employing a particular type of communication game referred to as parity oblivious communication game, it has been shown \cite{Pan2021} that the preparation noncontextual bound is $\mathcal{I}_{pnc}\leq 4$, here we revisit the evaluation of preparation noncontextual bound without invoking such game. 

Before proceeding further, let us briefly recapitulate the notion of preparation non-contextuality in the  CHSH scenario \cite{Clauser1969} where Alice (Bob) measures two observables $A_{1}$ and $A_{2}$ ($B_1$ and $B_2$). Alice's two measurements on her local part of the shared state $\rho_{AB}$ yield two density matrices on Bob's wing denoted as $\rho_{A_{1}}$ and $\rho_{A_{2}}$. Naturally,  $\rho_{A_{1}}=\rho_{A_{2}}\equiv \sigma$, else no-signalling condition will be violated. Such a feature can be assumed to be equivalent as represented in an ontological model, if one assumes that the ontological model is preparation non-contextual, i.e., $\mu(\lambda|\sigma, A_{1})=\mu(\lambda|\sigma, A_{2})$. It is then intuitively straightforward to conclude that a quantum violation of the CHSH inequality can also be regarded as proof of preparation contextuality. As demonstrated in \cite{Uola2020}, in the CHSH scenario, preparation non-contextuality implies the locality assumption. A modified version of this proof is outlined in \cite{Roy2023}.

For this, by using Bayes' theorem, the joint probability distribution in the ontological model can be expressed as follows \cite{Tavakoli2020jan, Roy2023} 
\begin{equation}
\label{jp}
 p(a, b|A_{x}, B_{y}) = \sum_{\lambda}p(a|A_{x}, B_{y}) p(\lambda|a,A_{x}) p(b|B_{y}, \lambda)
\end{equation}
Now, the locality condition implies the marginal probability of Alice's side is independent of Bob's choice of observables and hence we can write
\begin{equation}
\label{jpd}
 p(a, b|A_{x}, B_{y}) = \sum_{\lambda}p(a|A_{x})p(\lambda|a,A_{x})p(b|B_{y}, \lambda)
\end{equation}
From Bayes' theorem, it follows that $p(a|A_{x}) p(\lambda|a,A_{x})=\mu(\lambda|A_{x}) p(a|\lambda,A_{x})$ where we specifically denoted the probability distribution $p(\lambda|A_{x})$ as $\mu(\lambda|A_{x})$. Substituting this into Eq.~(\ref{jpd}), we get
\begin{equation}\label{aa}
 p(a, b|A_{x}, B_{y}) = \sum_{\lambda}\mu(\lambda|A_{x})p(a|\lambda,A_{x})p(b|B_{y}, \lambda)
\end{equation}

Assuming the preparation non-contextual for Bob's preparation by Alice, i.e., $\mu(\lambda|\rho,{A_{1}})=\mu(\lambda|\rho,{A_{2}})\equiv\mu(\lambda)$, we simplify Eq.~(\ref{aa}) to 
\begin{equation}
\label{}
 p(a, b|A_{x}, B_{y}) = \sum_{\lambda}\mu(\lambda)p(a|\lambda,A_{x})p(b| B_{y}, \lambda)
\end{equation}

This result aligns with the desired factorisability condition commonly derived for a local hidden variable model. Therefore, we argue that whenever joint probability distribution $p(a, b|A_{x}, B_{y})$ in the ontological model satisfies the assumption of preparation non-contextuality, it inherently satisfies locality condition in the ontological model.

Moving beyond the CHSH scenario, we introduce a nontrivial form of preparation non-contextuality in a Bell experiment involving more than two inputs. This involves imposing an additional relational constraint on Alice's measurement observables, expressed by the following condition
 \begin{equation}\label{pp1}
     \forall b, y; \sum_{x}P(a, b|A_{x},B_{y})=\sum_{x}P(a\oplus 1, b|A_{x},B_{y})
 \end{equation}
 
In quantum theory, when Alice and Bob share an entangled state $\rho_{AB}$, the above condition translates to 
 \begin{equation}\label{p34}
 \sum_{x}\rho^{a}_{A_{x}}=\sum_{x}\rho^{a\oplus 1}_{A_{x}}\equiv \sigma
 \end{equation}
 
 Here, $\rho^{a}_{A_{x}}=\Tr_{A}\left[\rho_{AB}\Pi^{a}_{A_{x}}\otimes \openone\right]$ and Eq.~({\ref{p34}}) implies $A_{1}+A_{2}+A_{3}=0$. This introduces a nontrivial constraint on Alice's preparation procedures, leading to a local bound reduced to the preparation non-contextual bound $\mathcal{I}_{pnc}\leq 4$.

Thus, nontrivial preparation contextuality provides a weaker notion of non-locality. It is important to note that the set of observables of Alice for which the optimal quantum value $\mathscr{I}_{opt}=6$ is achieved satisfies the condition given by Eq.~(\ref{pp1}). More precisely, our demonstration reveals that attaining the optimal quantum value $\mathscr{I}_{opt}=6$ necessitates the fulfilment of the constraint $A_{1}+A_{2}+A_{3}=0$.

\section{Optimal Quantum bound of the Bell functional $\mathcal{I}$} \label{obbf3}

We derive the optimal quantum value of the Bell functional, given by Eq.~(\ref{nbell}), devoid of assuming the dimension of the quantum system and by utilizing an elegant SOS approach introduced in \cite{Pan2020}. For this, we consider a positive semi-definite operator $\gamma$, which can be expressed as $\gamma=(\beta \ \openone_{d}-\mathcal{I})$, where $\beta$ is a positive real number and $\openone_{d}$ is the identity operator in an arbitrary $d$-dimensional system. This can be proven by considering a set of operators $L_{y}$, which are linear functions of the observables (Hermitian operators) $A_{x}$ and $B_{y}$, such that  
\begin{equation}\label{gamma}
		\gamma=\frac{1}{2}\sum_{y=1}^3 \omega_y \ L_y^\dagger L_y
\end{equation}
The operators $L_{y}$ are defined as follows 
	\begin{eqnarray}
		&&L_{y}=  \mathscr{A}_y\otimes \openone - \openone \otimes B_y \ \ \ \forall y \in [1,3] \label{li}
	\end{eqnarray}
 where $\mathscr{A}_y$ and $\omega_y$ defined as
\begin{eqnarray}
     &&\mathscr{A}_1 = \frac{A_1+A_2-A_3}{\omega_1} \ ; \ \ \mathscr{A}_2 = \frac{A_1-A_2+A_3}{\omega_2} \ ; \nonumber \\
    && \mathscr{A}_3 = \frac{-A_1+A_2+A_3}{\omega_3} \ ; \ \ \omega_y=||\mathscr{A}_y||_{\rho} \label{omegai}
\end{eqnarray}
where, $||\cdot||$ is the Frobenious norm, given by $||\mathcal{O}||=\sqrt{\Tr[\mathcal{O}^{\dagger}\mathcal{O} \ \rho]}$ and $\mathscr{I}=\Tr[\mathcal{I} \ \rho]$.

Now, putting Eqs.~(\ref{li}) and (\ref{omegai}) into Eq.~(\ref{gamma}) and by noting that $A_{x}^{\dagger} A_{x}=B_{y}^{\dagger} B_{y}=\openone_d$, we obtain
	\begin{eqnarray} \label{opt1}
		\Tr[\gamma \ \rho]&=&-\mathscr{I} + \sum_{y=1}^3 \omega_y
	\end{eqnarray}
	
	Therefore, it follows from the above Eq.~(\ref{opt1}) that the quantum optimal value of $\mathcal{I}$ is attained when $\langle\gamma\rangle=\Tr[\gamma\,\rho] = 0$, which in turn provides 
\begin{equation}\label{optbn}
\mathscr{I}_{opt}= \max\left(\sum\limits_{y=1}^{3}\omega_{y}\right)
\end{equation}

Evaluating $\omega_y$ from Eq.~(\ref{omegai}), we arrive at the following relations
\begin{eqnarray} 
\omega_{1}&=& \sqrt{3  + \Big\langle\{A_1 , A_2 - A_3\}\Big\rangle_\rho - \Big\langle\{A_2,A_3\}\Big\rangle_\rho} \nonumber \\ 
\omega_{2}&=& \sqrt{3  + \Big\langle\{A_1 , -A_2 + A_3\}\Big\rangle_\rho - \Big\langle\{A_2,A_3\}\Big\rangle_\rho} \nonumber \\ 
\omega_{3}&=& \sqrt{3  - \Big\langle\{A_1 , A_2 + A_3\}\Big\rangle_\rho +\Big \langle\{A_2,A_3\}\Big\rangle_\rho} \label{omegai1}
\end{eqnarray}

Next, using the convex inequality$\sum\limits_{i=1}^{n}\omega_{i}\leq \sqrt{n \sum\limits_{i=1}^{n} (\omega_{i})^{2}}$, from Eqs.~(\ref{optbn}) and (\ref{omegai1}), we get
\begin{eqnarray}\label{A1}
\mathscr{I}_{opt}&=& \max \sqrt{3 \ \qty(\omega_1^2+\omega_2^2+\omega_3^2)} \nonumber \\
&=&\max \sqrt{3 \ \qty[12-(A_1+A_2+A_3)(A_1+A_2+A_3)^\dagger]}\nonumber\\
&=& 6 \label{optbn1}
\end{eqnarray}

Clearly, the optimal value occurs when $A_1+A_2+A_3=0$, implying $\omega_1=\omega_2=\omega_3=2$. It is straightforward to check that $A_1+A_2+A_3=0$ implies $\Tr[\{A_x,A_{x'}\} \ \rho] = -1  \ \forall x\neq x' \in \{1,2,3\}$, consequently $\Tr[\{\mathscr{A}_y,\mathscr{A}_{y'}\} \ \rho]=-1  \ \forall y\neq y' \in \{1,2,3\}$. 

\subsection{The state and observables for achieving optimal violation}

The optimality condition $\Tr[\gamma \rho] = 0 \ \forall \rho$ implies that $\sum_{y=1}^{3} \Tr[L_y^\dagger L_y\ \rho] = 0 \ \forall \rho$. Given that $L_y^\dagger L_y$ are positive and Hermitian operators, this relation leads us to the crucial deduction that $\forall y,\Tr[L_y \ \rho] = 0$. Hence, Eq.~(\ref{li}) allows us to express the following
\begin{eqnarray}\label{Lic}
\Tr[L_{y} \ \rho]&=&0 \implies \Tr[\openone\otimes B_{y} 
 \ \rho]=\Tr[\mathscr{A}_y\otimes \openone  \ \rho]
\end{eqnarray}

Furthermore, for achieving optimal quantum violation, it is crucial for Alice's observable to satisfy the condition $\langle\{A_x, A_{x'}\}_{x\neq x'}\rangle_{\rho}=-1 \ \forall \rho$. As a consequence, $\langle\{\mathscr{A}_y, \mathscr{A}_{y'}\}_{y\neq y'}\rangle_{\rho}=-1 \ \forall \rho$. Therefore, based on Eq.~(\ref{Lic}), Bob's observable must satisfy $\forall \rho,\langle\{B_y, B_{y'}\}_{y\neq y'}\rangle_{\rho}=-1 \ \forall \rho$ for optimal violation.

Since the optimality condition is $\Tr[L_y^{\dagger}L_y \ \rho] =0$ and consequently Eq.~(\ref{li}) leads to the inference that $\Tr[\mathscr{A}_y\otimes B_y \ \rho]=1$ $\forall y$. Thus, $\rho$ must be a common eigenstate of the operators $\qty(\mathscr{A}_y\otimes B_y)$ with $y=1,2,3$. As these operators yield the maximum eigenvalue with the normalised state $\rho$, we can deduce that the state $\rho$ is pure. As a result, we can expand $\rho$ in terms of mutually commuting elements. However, since $\qty[\qty(\mathscr{A}_y\otimes B_y),\qty(\mathscr{A}_{y'}\otimes B_{y'})]_{y\neq y'}\neq 0$, we are unable to express $\rho$ in terms of $\qty(\mathscr{A}_y\otimes B_y)$ in contrast to the CHSH scenario (see Eq.~(7) of \cite{Roy2023}).

Let us first consider the general form of $\rho$ in the bipartite two-qubit scenario \cite{Horodecki1995}
\begin{eqnarray}
    \rho = \frac{1}{4}\qty[\openone\otimes\openone+\sum_{i=1}^3 r_i \ \sigma_i \otimes \openone+\sum_{j=1}^3 s_j \  \openone\otimes \sigma_j+\sum_{i,j=1}^3 t_{ij} \ \sigma_i\otimes \sigma_j]\nonumber\\
\end{eqnarray}
where $t_{ij}=\Tr[\sigma_i \otimes \sigma_j \ \rho]$ are the elements of the correlation matrix $T=[t_{ij}]$, $r_i$ and $s_j$ are positive real numbers, satisfying the following relations \cite{Horodecki1995}
\begin{equation} \label{statecondqubit}
\sum_{i=1}^3 r_{i}^2+\sum_{j=1}^3 s_{j}^2+\sum_{i,j=1}^3t_{ij}^2 \leq 3
\end{equation}
and equality in the above Eq.~(\ref{statecondqubit}) holds for pure states. 

Now, let us consider, there exists a two-qubit state $\rho$ and a set of Hermitian operators $\{C_i \neq \openone\} \ \forall i \in \{1,2,3\}$ such that the condition $\Tr[C_i \otimes C_i \rho]=1$ holds, then it is evident that $t_{ij}=0 \ \forall i\neq j$ and $t_{ij}=\pm 1 \ \forall i=j$, implying $\sum_{i,j=1}^3t_{ij}^2 = 3$. Consequently, from Eq.~(\ref{statecondqubit}), it follows that $\sum_{i=1}^3 r_{i}^2=\sum_{j=1}^3 s_{j}^2=0$. 

Therefore, for the condition $\Tr[C_i \otimes C_i \rho]=1$ to be satisfied, the bipartite qubit state must be of the following form
\begin{equation} \label{bqme}
     \rho = \frac{1}{4}\qty[\openone\otimes\openone+\sum_{i=1}^3 C_i \otimes C_i]
\end{equation}
where $C_i \otimes C_i=t_{ii} \ \sigma_i\otimes \sigma_i$, satisfying $[C_i \otimes C_i, C_j \otimes C_j]_{i\neq j}=0$. It is important to emphasise here that the state expressed by Eq.~(\ref{bqme}) is an entangled state, meeting the criterion $\Tr_A[\rho]=\Tr_B[\rho]=\frac{\openone}{2}$ and $\rho^2=\rho$. This signifies that $\rho$ is a pure maximally entangled two-qubit state. The significant insight gained here is that a maximally entangled state can be represented in terms of mutually commuting operators.

Expanding on this notion, if we extend the idea to express a maximally entangled state in any arbitrary dimension $d$, we can represent $\rho$ in terms of mutually commuting operators, $C_i\otimes C_i$, which are functions of both $\mathscr{A}_y$ and $B_y$. Thus, $\rho$ takes the form
\begin{eqnarray}\label{rhod}
\rho = \frac{1}{d^2} \qty[\openone_d\otimes\openone_d + \sum_{i=1}^{d^2-1} C_i \otimes C_i]
\end{eqnarray}
Now, any three $C_i \otimes C_i$ can be derived based on the following optimality conditions obtained from the SOS approach (i) $\Tr[\mathscr{A}_y\otimes B_y \ \rho]=1$, (ii) $\Tr[C_i \otimes C_i \rho]=1$, and (iii) $[C_i\otimes C_i, C_j\otimes C_j]_{i\neq j}=0$. Using such conditions, we obtain the following expressions for any three $C_i \otimes C_i$ (see Appendix~\ref{Co} for detailed derivations)
\begin{eqnarray}\label{Cs}
C_1 \otimes C_1 &=& \mathscr{A}_1 \otimes B_1 \nonumber \\
C_2 \otimes C_2&=&\frac{1}{3}\qty(\mathscr{A}_2 \otimes B_2 + \mathscr{A}_3 \otimes B_3 - \mathscr{A}_3 \otimes B_2 - \mathscr{A}_2 \otimes B_3) \nonumber \\
C_3 \otimes C_3 &=& \frac{1}{3}(\mathscr{A}_2 \mathscr{A}_1 \otimes B_2 B_1 - \mathscr{A}_2 \mathscr{A}_1 \otimes B_3 B_1 - \mathscr{A}_3 \mathscr{A}_1 \otimes B_2 B_1 \nonumber \\
&& + \mathscr{A}_3 \mathscr{A}_1 \otimes B_3 B_1)
\end{eqnarray}
Where $C_3 \otimes C_3$ satisfies the condition  $C_3 \otimes C_3= \qty(C_2 \otimes C_2)\cdot\qty(C_1 \otimes C_1)$.

An explicit example for two-qubit maximally entangled state is the following
\begin{equation}
\rho= \frac{1}{4}\qty(\openone\otimes \openone + \sigma_x \otimes \sigma_x - \sigma_y \otimes \sigma_y + \sigma_z \otimes \sigma_z)
\end{equation}

If we compare the given form of $\rho$ with the form given by Eq.~(\ref{rhod}), we can suitably choose $C_1\otimes C_1 = \sigma_x \otimes \sigma_x$, $C_2\otimes C_2 = \sigma_z \otimes \sigma_z$ and $ C_3\otimes C_3 =- \sigma_y \otimes \sigma_y $. Then, using Eqs.~(\ref{omegai}) and (\ref{Cs}), we determine the following particular set of observables that leads to the optimal quantum violation.
\begin{eqnarray}
  &&  A_1 = -B_3 = \frac{\sigma_x+\sqrt{3}\sigma_z}{2} \ ; \ A_2 = -B_2 = \frac{\sigma_x-\sqrt{3}\sigma_z}{2} \ ; \nonumber \\
  && \ A_3 = -B_1 = -\sigma_x  \ ; 
\end{eqnarray}
%%%%%%%%%%%%%%%%%%%%%%%%%%%%%%%%%%%%%%%%%%%%%%%%%%%%%%%%%%%%%%%%%%%%%%%%%%%%%%%%%%%%%%%%%%%%%%%%%%%%%%%%%%%%%%%%%%%%%%%%%%%%

\section{Sequential quantum violations of non-contextual inequality}\label{SSS}

\begin{figure*}
    \centering
    \includegraphics[width=\textwidth, height=6cm]{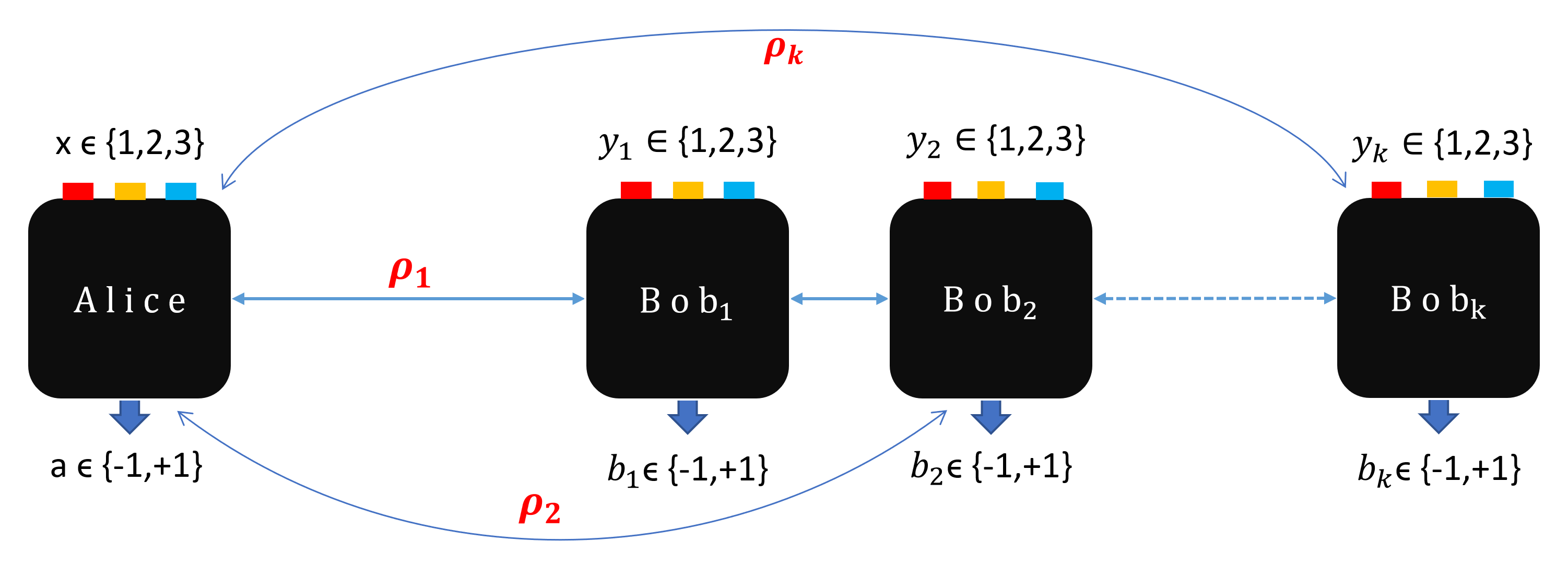}
    \caption{The diagram depicts the sequential Bell scenario. Alice and Bob$^{1}$ share an entangled state $\rho_{1} \in \mathscr{H}^d\otimes \mathscr{H}^d$. Alice always performs projective measurements and  Bob$^{k}$ performs unbiased unsharp POVMs. }\label{figseq}
\end{figure*}

Let's take a quick overview of the sequential Bell experiment scenario \cite{Silva2015}. Unlike the traditional Bell scenario with spatially separated Alice and Bob, we have one Alice who is spatially separated from multiple Bobs (see Fig.~\ref{figseq}).  Initially, a bipartite entangled state $\rho_{1} \in \mathscr{H}^d\otimes \mathscr{H}^d$ is shared between Alice and Bob$^{1}$. Following the reception of their respective subsystems, Alice performs projective measurement, and Bob$^{1}$ performs unsharp measurements (unbiased POVM). Then, Bob$^{1}$ passes his residual subsystem to the next sequential independent observer Bob$^{2}$. This process continues, with Bob$^2$ relaying his measured subsystem to Bob$^{3}$, and so forth.

The observables of Alice and Bob$^{k}$ are given by
\begin{eqnarray}
A_{x} &\equiv& \left\{ 
A_{\pm|x} \ \Big| \ A_{\pm|x} = \frac{1}{2}\qty(\openone \pm A_x) \right\} \ \ \forall x\in \{1,2,3\} \nonumber\\
\mathcal{B}_y^k&=& \Bigg\{\mathcal{E}_{\pm|y}^k \ \big| \ \mathcal{E}_{\pm|y}^k =\frac{1}{2}\qty(\openone\pm\eta_k B_y^k)\Bigg\} \ \ \forall y\in \{1,2,3\}\label{m1}
\end{eqnarray}
where $\mathcal{E}_{\pm|y}^k$ are POVM elements of Bob$^k$'s measurement, satisfying $\mathcal{E}_{\pm|y}^k\geq0$ and $\mathcal{E}_{+|y}^k+\mathcal{E}_{-|y}^k=\openone$.

The post-measurement state $\rho_{k}$ after $(k-1)^{th}$ Bob's measurement is evaluated \cite{Busch1996book} as
\begin{eqnarray}
\rho_k &=& \frac{1}{3}\sum\limits_{y=1}^3 \sum_{b=\pm} \qty(\openone \otimes \sqrt{B^{(k-1)}_{b|y}} ) \  \rho_{A B^{k-1}} \ \qty(\openone \otimes \sqrt{B^{(k-1)}_{b|y}}) \nonumber \\
&=& \frac{1}{3} \sum_{b,y} \qty[\qty(\openone \otimes \mathcal{K}^{k-1}_{b|y}) \ \rho_{k-1} \ \qty(\openone \otimes \mathcal{K}^{k-1}_{b|y})] \nonumber \\
&=& \frac{1+\xi_{k-1}}{2} \rho_{k-1} \nonumber \\
&&+ \frac{1-\xi_{k-1}}{6} \sum_{y=1}^3 \qty(\openone\otimes B_y^{k-1}) \rho_{k-1} \qty(\openone\otimes B_y^{k-1}) \label{stategen2}
\end{eqnarray}
where $\xi_j=\sqrt{1-\eta_j^2}$ and
\begin{eqnarray}
\mathcal{K}^{k}_{\pm|y} &=&  \frac{1}{2}\qty(\sqrt{\frac{1+\eta_k}{2}}+\sqrt{\frac{1-\eta_k}{2}}) \ \openone  \nonumber\\
&& \pm \frac{1}{2}\qty(\sqrt{\frac{1+\eta_k}{2}}-\sqrt{\frac{1-\eta_k}{2}}) \ B_y^k
\end{eqnarray}

The quantum value of the Bell functional($\mathcal{I}$) between Alice and Bob$^k$, $\mathscr{I}^k=\Tr[\mathcal{I} \ \rho_{k}]$ is evaluated as (see Appendix~\ref{aabk})
\begin{equation} \label{Ikf}
 \mathscr{I}^k  =
    \begin{cases}
    \eta_1 \ \sum\limits_{y=1}^3 \omega_y \ \Tr[\mathscr{A}_y\otimes B_y \ \rho_{1}] =\eta_1 \ \mathscr{I}_{opt} & \ \text{if} \  k=1 \\
       \eta_k \  \sum\limits_{y=1}^3 \omega_y \ \tilde{\omega}_y^k \Tr[\mathscr{A}_y\otimes \mathscr{B}_y^k \ \rho_{k-1}] & \ \text{if} \ k\geq2
    \end{cases}       
\end{equation}
where $\mathscr{A}_y$ and $\omega_y$ are given by Eq.~(\ref{omegai}), $\mathscr{B}_y^k = \tilde{B_y}^k/\tilde{\omega}_y^k$ with $\tilde{\omega}_y^k=||\tilde{B}_y^k||_{\rho_{k-1}}$ and
\begin{eqnarray}
    \Tilde{B}_y^k = \frac{1+\xi_{k-1}}{2} \ B_y^k + \frac{1-\xi_{k-1}}{6} \ \sum\limits_{y'=1}^{3} B_{y'}^{k-1} B_y^k B_{y'}^{k-1} \ \ \ \forall y
\end{eqnarray}

Since $\mathscr{I}_{opt}= 6$, it follows from Eq.~(\ref{Ikf}) that the sub-optimal quantum Bell value between Alice-Bob$^1$ is $\mathscr{I}^1 = 6 \eta_1$.

\begin{Lemma}\label{lemm}
    The sub-optimal quantum Bell value between Alice-Bob$^1$ is $\mathscr{I}^1 = 6 \eta_1$ with Bob$^1$'s observable satisfy the condition $B_1^1+B_2^1+B_3^1=0$, thereby implying $\langle \{B_y^1,B_{y'}^1\}_{y \neq y'} \rangle_{\rho_1} = -1$.
\end{Lemma}

We prove the following theorems.
\begin{thm} \label{th1}
     If Alice and Bob$^1$ obtain the sub-optimal Bell value of $\mathscr{I}^1=6\eta_1$, the sub-optimal quantum Bell value between Alice and Bob$^2$ is $\mathscr{I}^2 = 3 \ \eta_2 \qty(1+\sqrt{1-\eta_1^2})$ with both Bob's observables satisfy the condition, $B_1^k+B_2^k+B_3^k=0$, thereby implying $\langle\{B_y^k,B_{y'}^k\}_{y\neq y'}\rangle_{\rho_1}=-1$  with $k\in\{1,2\}$. 
\end{thm}

\begin{proof}
From Eq.~(\ref{Ikf}) $\mathscr{I}^2$ is
\begin{eqnarray}
    \mathscr{I}^2 = \eta_2 \  \sum\limits_{y=1}^3 \omega_y \ \tilde{\omega}_y^2 \Tr[\mathscr{A}_y\otimes \mathscr{B}_y^2 \ \rho_{1}]
\end{eqnarray}

where $\mathscr{B}_y^2 = \tilde{B}_y^2/\tilde{\omega}_y^2$, $\tilde{\omega}_y^2=||\tilde{B}_y^2||_{\rho_{1}}$ and
\begin{eqnarray}\label{By2g}
     \Tilde{B}_y^2 = \frac{1+\xi_1}{2} \ B_y^2 + \frac{1-\xi_1}{6} \ \sum\limits_{y'=1}^{3} B_{y'}^{1} B_y^2 B_{y'}^{1}  \ \forall y
\end{eqnarray}

For optimizing $\mathscr{I}^2$, we again employ the SOS approach and define a positive operator as follows
\begin{equation}\label{gammap}
		\gamma=\frac{1}{2}\sum_{y=1}^3 \omega_y\Tilde{\omega}_y^2 L_y^\dagger L_y
\end{equation}
The operators $L_i$ are given by
\begin{eqnarray}
		&&L_{y}=  \mathscr{A}_y\otimes \openone - \openone \otimes \frac{\Tilde{B}_y^2 }{\Tilde{\omega}_y^2} \ ; \ \Tilde{\omega}_y^2 = ||\Tilde{B}_y^2||_{\rho_1} \ \forall y \in \{1,2,3\} \label{omab2p}
	\end{eqnarray}
Now, similar to the method presented in Sec.~(\ref{obbf3}), from Eqs.~(\ref{gammap}) and (\ref{omab2p}), we obtain 
 \begin{equation}
\mathscr{I}^2 = \eta_2 \ \qty[\max \sum\limits_{y=1}^{3} \omega_y \  \Tilde{\omega}_y^2] \ \ \ \ (\because \Tr[\gamma\rho_1]=0)
 \end{equation}
Next, invoking the inequality $\sum\limits_y \sqrt{r_y s_y}\leq \sqrt{\sum\limits_y r_y}\sqrt{\sum\limits_y s_y}$, we get
\begin{eqnarray}\label{I2fopt}
\mathscr{I}^2 &\leq& \eta_2 \max \sqrt{(\omega_1)^2+(\omega_2)^2+(\omega_3)^2} \nonumber \\
&&\times \max\sqrt{(\Tilde{\omega}_1^2)^2+(\Tilde{\omega}_2^2)^2+(\Tilde{\omega}_3^2)^2}
 \end{eqnarray}
Note that equality holds when $\omega_y=\omega_{y'}$ and  $\Tilde{\omega}_y = \Tilde{\omega}_{y'}$. From Eq.~(\ref{omegai1}), we already obtain $\max \sqrt{\omega_1^2+\omega_2^2+\omega_3^2} = 2\sqrt{3}$ when $\langle \{A_y,A_{y'}\}_{y\neq y'}\rangle_{\rho_1}=-1$ and $A_1+A_2+A_3=0$. The optimality condition $\Tr[\gamma \rho_1]=0$ implies that $\Tr[L_y \ \rho_1]=0$. This means that $\mathscr{A}_y \otimes \openone=\openone\otimes \frac{\Tilde{B}_y^2}{\Tilde{\omega}_y^2}$. Therefore, $A_1+A_2+A_3=0$ implies $\Tilde{B}_1^2+\Tilde{B}_2^2+\Tilde{B}_3^2=0$. By using this relation, we obtain (see Appendix~\ref{aab2c})
\begin{eqnarray}\label{c1}
   \Big\langle \qty{B_y^2,B_{y'}^2}_{y \neq y'} \Big\rangle_{\rho_1} = -1 \ \ \ \forall y,y \in \{1,2,3\}
\end{eqnarray}
and
\begin{eqnarray}\label{c2}
   \sum_{y,y'=1}^3 B_{y'}^1 B_y^2 B_{y'}^1 = 0
\end{eqnarray}

Now, by using the relations $\tilde{\omega}_y^2 = \tilde{\omega}_{y'}^2$ and Eq.~(\ref{c2}), we get (see Appendix~\ref{aab2c1})
\begin{eqnarray}
 \sum_{y'=1}^3 B_{y'}^1 B_y^2 B_{y'}^1 &=& 0  \ \ \forall y \in \{1,2,3\} \label{c3}
 \end{eqnarray}
Using Eq.~(\ref{c3}) in Eq.~(\ref{By2g}), we arrive
\begin{eqnarray}\label{By2}
     \Tilde{B}_y^2 = \frac{1+\xi_1}{2} \ B_y^2
\end{eqnarray}

It is evident from Eq.~(\ref{By2}) that
\begin{eqnarray}\label{wk}
\Tilde{\omega}_y^2=||\Tilde{B}_y^2||_{\rho_1}=\frac{1+\xi_1}{2} = \frac{1+\sqrt{1-\eta_1^2}}{2}
\end{eqnarray}
Using Eq.~(\ref{wk}), the sub-optimal quantum Bell value between Alice-Bob$^2$ is (from Eq.~(\ref{I2fopt}) as derived
\begin{eqnarray}
    \mathscr{I}^2 = 3 \eta_2 \ \qty(1+\sqrt{1-\eta_1^2})
\end{eqnarray}
\end{proof}

\begin{thm} \label{th2}
If Alice-Bob$^1$ and Alice-Bob$^2$ obtain the sub-optimal Bell violations, the sub-optimal quantum Bell value between Alice and Bob$^3$ is $\mathscr{I}^3 = \frac{3 \ \eta_3}{2} \qty(1+\sqrt{1-\eta_1^2})\qty(1+\sqrt{1-\eta_2^2})$ with Bob$^k$'s observables satisfy the condition, $B_1^k+B_2^k+B_3^k =0$, thereby implying $\langle\{B_y^k, B_{y'}^k\}_{y\neq y'}\rangle_{\rho_1}=-1$ with $k\in\{1,2,3\}$.  
\end{thm}

\begin{proof}

By using Eq.~(\ref{stategen2}), we can write $\rho_2$ in terms of $\rho_1$ and from Eq.~(\ref{Ikf}) we obtain the value of $\mathscr{I}^3$ in terms of $\rho_1$ as follows

\begin{eqnarray}
    \mathscr{I}^3 = \eta_3 \  \sum\limits_{y=1}^3 \omega_y \ \tilde{\omega}_y^3 \Tr[\mathscr{A}_y\otimes \mathscr{B}_y^3 \ \rho_{1}]
\end{eqnarray}
 where $\mathscr{B}_y^3 = \frac{\tilde{B}_y^3}{\tilde{\omega}_y^3}$, $\tilde{\omega}_y^3=||\tilde{B}_y^3||_{\rho_{1}}$ and
 \begin{eqnarray}
     \Tilde{B}_y^3 &=& \frac{(1+\xi_{1})(1+\xi_{2})}{4} \ B_y^3 + \frac{(1-\xi_{1})(1+\xi_{2})}{12} \sum\limits_{{y'}=1}^{3} B_{y'}^{1} B_y^3 B_{y'}^{1}\nonumber\\
     &+&\frac{(1+\xi_{1})(1-\xi_{2})}{12} \sum\limits_{{y'}=1}^{3} B_{y'}^{2} B_y^3 B_{y'}^{2}+\frac{(1-\xi_{1})(1-\xi_{2})}{36} \nonumber\\
     &\times& \sum_{y', y''= 1}^3  B_{y''}^1 B_{y'}^2 B_y^3 B_{y'}^2 B_{y''}^1
 \end{eqnarray}
 
 From Lemma \ref{lemm} and Theorem \ref{th1}, we can see that Bob$^1$'s measurement settings and Bob$^2$'s measurement settings satisfy the same condition i.e, $(B_1^k+B_2^k+B_3^k)_{k\in\qty{1,2}}=0$. Then, without loss of generality, we can take $B_y^2=B_y^1  \ \ \forall y\in \qty{1,2,3}$. With this condition $\Tilde{B}_y^3$ simplified as as
 \begin{eqnarray}\label{B3f}
     \Tilde{B}_y^3 &=& \frac{(1+\xi_{1})(1+\xi_{2})}{4} \ B_y^3 + \frac{(1-\xi_{1}\xi_{2})}{6}\sum\limits_{{y'}=1}^{3} B_{y'}^{1} B_y^3 B_{y'}^{1}\nonumber\\&+&\frac{(1-\xi_{1})(1-\xi_{2})}{36} \sum_{{y',y''}=1}^3 B_{y''}^1 B_{y'}^1 B_y^3 B_{y'}^1 B_{y''}^1
 \end{eqnarray}

For optimising $\mathscr{I}^3$, we follow the approach similar to the proof of Theorem \ref{th1}. We define a positive operator
\begin{equation}\label{gammap1}
		\gamma=\frac{1}{2}\sum_{y=1}^3 \omega_y\Tilde{\omega_y}^3 L_y^\dagger L_y
\end{equation}
The operators $L_y$ are defined as follows
\begin{eqnarray}
L_{y}=  \mathscr{A}_y\otimes \openone - \openone \otimes \frac{\Tilde{B}_y^3 }{\Tilde{\omega}_y^3} \ ;  \ \ \Tilde{\omega}_y^3 = ||\Tilde{B}_y^3||_{\rho_1} \ \forall y \in \{1,2,3\} \label{omab2p1}
\end{eqnarray}

Then, similar to the argument presented in Theorem \ref{th1}, we obtain
\begin{eqnarray}\label{I3f}
    \mathscr{I}^3 &\leq& \eta_3 \max \sqrt{\omega_1^2+\omega_2^2+\omega_3^2} \nonumber\\ &\times&\max\sqrt{(\Tilde{\omega}_1^3)^2+(\Tilde{\omega}_2^3)^2+(\Tilde{\omega}_3^3)^2}
\end{eqnarray}

Where the equality holds when $\omega_y=\omega_{y}$ and  $\Tilde{\omega}_y^3=\Tilde{\omega}_{y}^3$, leading to the condition $\Tilde{B}_1^3+\Tilde{B}_2^3+\Tilde{B}_3^3=0$. Invoking this condition, we obtain the following relations (see Appendix~(\ref{appI3})
\begin{eqnarray}
   &&\Big\langle \qty{B_y^3,B_{y'}^3}_{y\neq y'} \Big\rangle_{\rho_1} = -1 \ \ \ \forall y,y' \in \{1,2,3\} \label{co3} \\
   && \sum_{y,y'=1}^3 B_{y'}^{1} B_y^3 B_{y'}^{1}=\sum_{y,y',y''=1}^3 B_{y''}^1 B_{y'}^1 B_y^3 B_{y'}^1 B_{y''}^1=0   \label{thm21}
\end{eqnarray}

By, using the relations $\tilde{\omega}_y^3 = \tilde{\omega}_{y'}^3$ and Eq.~(\ref{thm21}), we get (see Appendix~\ref{appI4})
\begin{eqnarray}
 &&\sum_{y'=1}^3 B_{y'}^1 B_y^2 B_{y'}^1 = \sum_{y',y''= 1}^3 B_{y''}^1 B_{y'}^1 B_y^3 B_{y'}^1 B_{y''}^1=0  \label{thm24}
 \end{eqnarray}
Now, by inserting Eq.~(\ref{thm24}) in Eq.~(\ref{B3f}), we obtain
\begin{eqnarray}\label{B_y^3}
    \Tilde{B}_y^3 = \frac{(1+\xi_{1})(1+\xi_{2})}{4} \ B_y^3
\end{eqnarray}
Which in-turn gives
\begin{eqnarray}\label{omega3}
    \Tilde{\omega}_y^3 = ||\Tilde{B}_y^3||_{\rho_1} = \frac{(1+\xi_{1})(1+\xi_{2})}{4} \ \ \ \forall y\in \qty{1,2,3}
\end{eqnarray}

Therefore, from Eqs.~(\ref{I3f}) and ~(\ref{omega3}), the sub-optimal quantum Bell value between Alice-Bob$^3$ is given by
\begin{eqnarray}\label{I3f1}
    \mathscr{I}^3 =  \frac{3 \eta_3}{2} \qty(1+\sqrt{1-\eta_1^2})\qty(1+\sqrt{1-\eta_2^2})
\end{eqnarray}

\end{proof}

%%%%%%%%%%%%%%%%%%%%%%%%%%%%%%%%%%%%%%%%%%%%%%%%%%%%%%%%%%%%%%%%%%%%%%%%%%%%%%%%%%%%%%%%%%%%%%%%%%%%%%%%%%%%%%%%%%%%%%%%%%%%%%%%%%%%%%%%%%%%%%%%%%%%%%%%%%%%%%%%%%%%%%%%%%%%%%%%%%%%%%%%%%%%%%%%%%%%%%%%%%%%%%%%%%%%%%%%%%%%%%%%%%%%%%%%%%%%%%%%%%%%%%%%%%%%%%%%%%%%%%%%%%%%%%%%%%%%%%%%%%%%%%%%%%%%%%%%%%%%%%%%%%%%%%%%%%%%%%%%%%%%%%%%%%%%%%%%%%%%%%%%%%%%%%%%%%%%%%%%%%%%%%%%%%%%%%%%%%%%%%%%%%%%%%%%%%%%%%%%%%%%%%%%%%%%%%%%%%%%%%%%%%%%%%%%%%%%%%%%%%%%%%%%%%%%%%%%%%%%%%%%%%%%

\begin{thm} \label{th3}
If Alice-Bob$^1$, Alice-Bob$^2$ and Alice-Bob$^3$ obtain the sub-optimal Bell violations, then no further sequential Bob gets the Bell violation.
\end{thm}

\begin{proof}
    From the Lemma \ref{lemm} and the Theorems \ref{th1} and \ref{th2}, we prove that Alice-Bob$^1$, Alice-Bob$^2$, and Alice-Bob$^3$ will obtain sub-optimal violation if all Bob's observables satisfy the conditions $B^k_1+B^k_2+B^k_3=0$ and $\{B^k_i, B^k_j\}_{i\neq j} =-1$ for all $k\in\{1,2,3\}$. Now, without loss of generality, we can always take $B_i^k=B_j^k \ \ \forall k \in\{1,2,3\}$. By, using these facts, from Eq.~(\ref{Ikf}), we obtain
    \begin{eqnarray}
        \mathscr{I}^4 = \frac{3 \eta_4}{4} \prod\limits_{j=1}^3 \qty(1+\sqrt{1-\eta_j^2}) \label{chsh4}
    \end{eqnarray}

Notably, in the context of Alice-Bob$^1$ violation, the bound of $\eta_1$ satisfies $\frac{2}{3}<\eta_1\leq 1$. Additionally, it is evident that $\frac{2}{3}<\eta_1<\eta_2<\eta_3$. Thus, the maximum Bell value achievable by Alice-Bob$^4$ is given by
\begin{equation}\label{chsh41}
\mathscr{I}^4 = \frac{3}{4} \qty(1+\frac{\sqrt{5}}{3})^3 \approx 3.9876 <4 
\end{equation}
\end{proof}

To put a succinct discussion of the preceding Section \ref{SSS}, we have evaluated quantum sub-optimal Bell values for Alice-Bob$^1$ ($\mathscr{I}^1$), Alice-Bob$^2$ ($\mathscr{I}^2$), and Alice-Bob$^3$ ($\mathscr{I}^3$). Subsequent sections will elaborate on the way these concurrently sub-optimal values play a pivotal role in certifying the unsharp parameters under the condition $\{\mathscr{I}^1, \mathscr{I}^2, \mathscr{I}^3\} > 4$. 
%%%%%%%%%%%%%%%%%%%%%%%%%%%%%%%%%%%%%%%%%%%%%%%%%%%%%%%%%%%%%%%%%%%%%%%%%%%%%%%%%%%%%%%%%%%%%%%%%%%%%%%%%%%%%%%%%%%%%%%%%%%%%%%%%%%%%%%%%%%%%%%%%%%%%%%%%%%%%%%%%%%%%%%%%%%%%%%%%%%%%%%%%%%%%%%%%%%%%%%%%%%%%%%%%%%%%%%%%%%%%%%%%%%%%%%%%%%%%%%%%%%%%%%%%%%%%%%%%%%%%%%%%%%%%%%%%%%%%%%%%%%%%%%%%%%%%%%%%%%%%%%%%%%%%%%%%%%%%%%%%%%%%%%%%%%%%%%%%%%%%%%%%%%%%%%%%%%%%%%%%%%%%%%%%%%%%%%%%%%%%%%%%%%%%%%%%%%%%%%%%%%%%%%%%%%%%%%%%%%%%%%%%%%%%%%%%%%%%%%%%%%%%%%%%%%%%%%%%%%%%%%%%%%%

\section{Self-testing of  unsharp quantum instruments}\label{DIS}

Following Lemma.~\ref{lemm}, Theorems \ref{th1} and \ref{th2}, it is evident that Alice-Bob$^1$, Alice-Bob$^2$ and Alice-Bob$^3$ will attain simultaneous violations of the preparation noncontextual bound of the Bell inequality given by Eq.~(\ref{nbell}), if $\{\mathscr{I}^1, \mathscr{I}^2, \mathscr{I}^3\} > 4$. 

Now, from the Lemma~\ref{lemm}, it is deduced that
\begin{eqnarray}
 \mathscr{I}^1>4 \implies \frac{2}{3}<\eta_1\leq 1    
\end{eqnarray}

Likewise, from the Theorem~\ref{th1}, we obtain
\begin{eqnarray}\label{I2b}
\mathscr{I}^2>4 \implies \frac{4}{3\qty(1+\sqrt{1-\eta_1^2})}<\eta_2\leq1 \label{l2}
\end{eqnarray}

It's worth noting that since $0<\eta_2\leq 1$, Eq.~(\ref{l2}) establishes a lower bound for $\eta_2$, which consequently fixes the upper bound of $\eta_1$. Thus, we have $\frac{2}{3}< \eta_1 \leq \frac{2\sqrt2}{3}$.

Next, the Theorem~\ref{th2} reveals that the sub-optimal quantum Bell violation for Alice-Bob$^3$, i.e, $\mathscr{I}^3>4$, imposes constraints on both the upper bound of $\eta_2$ and the lower bound of $\eta_3$, evaluated as follows
\begin{equation}
\eta_2 <\frac{4 \sqrt{3\sqrt{1-\eta_1^2}-1}}{3 \ \qty(1+\sqrt{1-\eta_1^2})} \label{l3}
\end{equation}
\begin{equation}
\frac{8}{3 \ \qty(1+\sqrt{1-\eta_1^2})\qty(1+\sqrt{1-\eta_2^2})} < \eta_3 \leq 1 \label{lambda3b}
\end{equation}
Together with Eqs.~(\ref{l2}) and (\ref{l3}), we derive
\begin{equation}
    \frac{4}{3 \ \qty(1+\sqrt{1-\eta_1^2})}<\eta_2< \frac{4 \sqrt{3\sqrt{1-\eta_1^2}-1}}{3 \ \qty(1+\sqrt{1-\eta_1^2})} \label{l4}
\end{equation}
Notably, the upper and lower bounds of $\eta_2$ consequently restrict the upper bound of $\eta_1$, specifically $\eta_1<\frac{\sqrt{5}}{3}$. Therefore, the lower and upper bounds for $\eta_1$ are given by
\begin{eqnarray}
    \frac{2}{3}<\eta_1< \frac{\sqrt{5}}{3} \label{lambda2b}
\end{eqnarray}

Hence, for $\{\mathscr{I}^1,\mathscr{I}^2,\mathscr{I}^3\} > 4$, we find the bounds of three unsharpness parameters as given by Eqs.~(\ref{lambda3b})-(\ref{lambda2b}).

From the Lemma~\ref{lemm} and the Theorem~\ref{th1} we have $\mathscr{I}^1 = 6\eta_1$ and $\mathscr{I}^2 = 3 \ \eta_2 \qty(1+\sqrt{1-\eta_1^2})$. It is evident that an increase in the unsharpness parameter $\eta_1$ for Bob$^1$ leads to a decrease in the sub-optimal Bell value between Alice and Bob$^2$. This implies that the more Bob$^1$ extracts information, the less the value of $\mathscr{I}^2$ is obtained. Thus, there exists a trade-off between the sub-optimal quantum Bell values for Alice-Bob$^1$ and Alice-Bob$^2$. 

Additionally, by applying the Theorem~\ref{th2} and rewriting $\mathscr{I}^3$ in terms of $\mathscr{I}^1$ and $\mathscr{I}^2$, we get
\begin{widetext}
    \begin{eqnarray}\label{I3t}
    \mathscr{I}^3 &=& \frac{3}{2} \ \qty[1+\sqrt{1-\qty(\frac{\mathscr{I}^1}{6})^2}] \ \ \qty[1+\sqrt{1-\frac{\qty(\mathscr{I}^2)^2}{9 \ \qty{1+\sqrt{1-\qty(\frac{\mathscr{I}^1}{6})^2}}^2}}]
\end{eqnarray}
\end{widetext}

\begin{figure}[!ht] 
\includegraphics[scale=0.4]{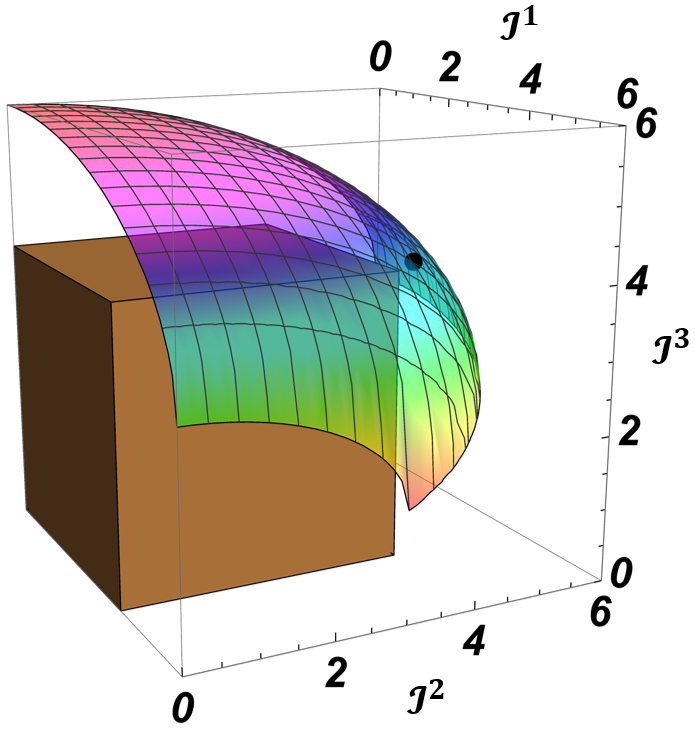}

\caption{The graph illustrates the trade-off among the sub-optimal quantum bounds for the Bell values between Alice-Bob$^1$, Alice-Bob$^2$ and Alice-Bob$^3$. The brown-coloured cuboid depicts the preparation noncontextual region of the concerned Bell inequality. The coloured curved surface indicates the region where the optimal quantum values of the Bell function exceed the preparation noncontextual bound, which is four. The black point on the three-dimensional graph holds significance as it serves as a crucial point certifying the unsharpness parameters $\eta_1$ and $\eta_2$ when the quantum values of all three sequential Bobs are considered to be equal. Each point on the curved surface represents a distinct set of sequential Bell values that self-tests a particular set of unsharpness parameters $\qty{\eta_1, \eta_2}$.}
\label{fig1}
\end{figure}

The optimal trade-off between the sub-optimal quantum bounds of the Bell inequality for Alice-Bob$^1$, Alice-Bob$^2$ and Alice-Bob$^3$ is given by Eq.~(\ref{I3t}) and illustrated in Fig.~(\ref{fig1}). It is important to highlight here that $\mathscr{I}^3>4$ implies $4<\mathscr{I}^1<2\sqrt{5}$ and $4<\mathscr{I}^2<4 \sqrt{\frac{1}{2}\sqrt{36-\qty(\mathscr{I}^1)^2}-1}$. The lower bound of $\mathscr{I}^1$ determines the upper bound of $\mathscr{I}^2$ i.e, $4<\mathscr{I}^2<4\sqrt{\sqrt{5}-1}$.

Now, given that $\frac{\mathscr{I}^1}{42}<<1$ and $\frac{\mathscr{I}^2}{42}<<1$, we can expand the right hand side of Eq.~(\ref{I3t}) using the Taylor series expansion. Furthermore, by neglecting higher order terms of $\frac{\mathscr{I}^1}{42}$ and $\frac{\mathscr{I}^2}{42}$-specifically, taking $\mathcal{O}\qty{\qty(\frac{\mathscr{I}^1}{42})^m}= \mathcal{O}\qty{\qty(\frac{\mathscr{I}^2}{42})^m}=0$ with $m\geq 3$ and $\mathcal{O}\qty{\qty(\frac{\mathscr{I}^1\mathscr{I}^2}{36})^n}=0$ with $n\geq 2$, we derive
\begin{equation}
    -\qty(\mathscr{I}^3- 6) \approx \frac{3}{2}\qty{\qty(\frac{\mathscr{I}^1}{6})^2 +  \ \qty(\frac{\mathscr{I}^2}{6})^2}
\end{equation}

This equation represents a paraboloid about negative $\mathscr{I}^3$ axis, with the origin shifted to $(6,0,0)$. We are particularly interested in the region where $\mathscr{I}^k \in [4,6]$ with $k=\{1,2,3\}$ and consider only the half-section of the paraboloid, i.e, semi-paraboloid\footnote{A semi-paraboloid is a three-dimensional geometric shape that is formed by taking a paraboloid and cutting it along a plane. It is essentially half of a paraboloid. The term ``semi" indicates that only half of the paraboloid is considered, resulting in a curved surface resembling a half-bowl or half-umbrella.}. 

It is crucial to emphasise that for sharp measurements of Bob$_3$, meaning $\eta_3 = 1$, each point on the surface of the semi-paraboloid in Fig.~(\ref{fig1}) uniquely certifies the pair $(\eta_1, \eta_2)$. For instance, the black point on the surface of the semi-paraboloid uniquely certifies $\qty(\eta_1=\frac{20}{29}\approx 0.69, \eta_2 = \frac{4}{5}\approx 0.8 )$ for $\mathscr{I}^1=\mathscr{I}^2=\mathscr{I}^3=\frac{120}{29} \approx 4.14$.

\subsection{Self-testing statement in sequential scenario}\label{cor1}
From the Lemma \ref{lemm}, Theorems \ref{th1} and \ref{th2}, the optimal tuple $\{\mathscr{I}^1,\mathscr{I}^2,\mathscr{I}^3\}$ uniquely certifies the state shared between Alice and Bob$^1$, along with the observables of both Alice or Bob and the unsharpness parameters of Bob. The self-testing statements are the following
\begin{enumerate}  [(i)]
\item The initial shared state between Alice and Bob$^1$ is a Bipartite maximally entangled state in any arbitrary dimension.
\item Alice performs projective measurements on her subsystem in any arbitrary local dimension. For the optimal values of the tuple, Alice's observables must satisfy the condition $A_1 + A_2 + A_3 =0$,  revealing a particular anti-commutation relation between her observables, given by $\{A_i, A_j\}_{i\neq j}=-1$.
\item Bob$^k$ performs unsharp measurements on his subsystem in any arbitrary local dimension, which satisfies the conditions $B_1^k+B_2^k+B_3^k=0$ and $\{B_i^k, B_j^k\}_{i\neq j}=-1$ for all $k\in\{1,2,3\}$.
\item The optimal tuple  $\{\mathscr{I}^1,\mathscr{I}^2,\mathscr{I}^3\}$ uniquely certifies the pair of unsharpness parameters $\{\eta_1,\eta_2\}$. In other words, the specific values of $\mathscr{I}^1$, $\mathscr{I}^2$, and $\mathscr{I}^3$ allow for a precise determination of the unsharpness parameters $\eta_1$ and $\eta_2$ associated with the experiment.
\end{enumerate}

%%%%%%%%%%%%%%%%%%%%%%%%%%%%%%%%%%%%%%%%%%%%%%%%%%%%%%%%%%%%%%%%%%%%%%%%%%%%%%%%%%%%%%%%%%%%%%%%%%%%%%%%%%%%%%%%%%%%%%%%%%%%%%%%%%%%%%%%%%%%%%%%%%%%%%%%%%%%%%%%%%%%%%%%%%%%%%%%%%%%%%%%%%%%%%%%%%%%%%%%%%%%%%%%%%%%%%%%%%%%%%%%%%%%%%%%%%%%%%%%%%%%%%%%%%%%%%%%%%%%%%%%%%%%%%%%%%%%%%%%%%%%%%%%%%%%%%%%%%%%%%%%%%%%%%%%%%%%%%%%%%%%%%%%%%%%%%%%%%%%%%%%%%%%%%%%%%%%%%%%%%%%%%%%%%%%%%%%%%%%%%%%%%%%%%%%%%%%%%%%%%%%%%%%%%%%%%%%%%%%%%%%%%%%%%%%%%%%%%%%%%%%%%%%%%%%%%%%%%%%%%%%%%%%

\section{Robust certification of Unsharpness parameter}\label{rc}

Since the experimental demonstration of achieving the tuple $\{\mathscr{I}^1,\mathscr{I}^2,\mathscr{I}^3\}$ is always subject to unavoidable noises and imperfections, here, we investigate to what extent the self-testing statements, presented in Sec. (\ref{cor1}), remain valid and reliable in presence of such noises. Specifically, when the optimal tuple $\{\mathscr{I}^1,\mathscr{I}^2,\mathscr{I}^3\}$ cannot be realised, it becomes impossible to uniquely certify the pair $\{\eta_1, \eta_2\}$. Instead, in such instances, it is only possible to certify the ranges of the pair  $\{\eta_1, \eta_2\}$.

It is important to note that according to the Lemma~\ref{lemm}, Alice-Bob$^1$ obtains Bell violation when $\eta_1$ lies in the range $(\frac{2}{3},1]$. Consequently, if $\mathscr{I}^1 > 4$, it implies that $(\eta_1)_{min} = \frac{2}{3}$. However, the range of $\eta_1$ that can be certified will become more restricted when nonlocality is extended further between Alice and Bob$^2$. 

Now, for sequential Bell violation between Alice-Bob$^1$ and Alice-Bob$^2$, i.e., when $\mathscr{I}^1$ and $\mathscr{I}^2$ both exceed $4$, we employ the Theorem~\ref{th1} to determine an upper bound of $\eta_1$. In this case, with $\eta_2 = 1$, we find that $(\eta_1)_{max}=\frac{2\sqrt{2}}{3}$. Therefore, the simultaneous Bell violation between Alice-Bob$^1$ and Alice-Bob$^2$ certifies a range of $\eta_1$, specifically $\frac{2}{3}<\eta_1<\frac{2\sqrt2}{3}$.

Furthermore, if all three Bobs demonstrate Bell violations with Alice independently, it further narrows down the range of $\eta_1$. As a result, the allowed interval for $\eta_1$ becomes even more restricted compared to the previous case. In order to evaluate this range, we utilise Eq.~(\ref{lambda2b}), which yields the interval $\eta_1\in(\frac{2}{3},\frac{\sqrt{5}}{3})$. In addition to this, the presence of Bell violation between Alice-Bob$^3$ implies a range for $\eta_2$, which is given by $\eta_2\in\qty(3-\sqrt{5},\frac{4}{5})$. 

Therefore, when Alice-Bob$^1$, Alice-Bob$^2$ and Alice-Bob$^3$ all demonstrate preparation contextuality bound of the concerned Bell inequality, a particular range for both $\eta_1$ and $\eta_2$ are certified. In conclusion, if $\mathscr{I}^1,\mathscr{I}^2,\mathscr{I}^3>4$, then it is established that $\eta_1$ lies in the interval $\qty(\frac{2}{3},\frac{\sqrt{5}}{3})$ and $\eta_2$ lies in the interval $\qty(3-\sqrt{5},\frac{4}{5})$. 

Finally, by considering the given expression in Eq.~(\ref{I3f1}), we find that $\mathscr{I}^3>4$ fixes the lower bound of $\eta_{3}$, which is given by
\begin{eqnarray}
   (\eta_3)_{min}&=&\frac{ 2 \ \mathscr{I}^3}{ 3 \ \qty(1+\sqrt{1-(\eta_1)_{min}^2})\qty(1+\sqrt{1-(\eta_2)_{min}^2})} \nonumber \\
   &=&  \frac{1}{2} \qty(3+\sqrt{5}-\sqrt{6\sqrt{5}-2}) \approx 0.93
\end{eqnarray}

It is essential to underscore that the power of the sequential scenario compared to the usual Bell scenario, not only facilitate the certification of the unsharpness parameters but also enable the certification of the incompatibility of sequential measurements.
%%%%%%%%%%%%%%%%%%%%%%%%%%%%%%%%%%%%%%%%%%%%%%%%%%%%%%%%%%%%%%%%%%%%%%%%%%%%%%%%%%%%%%%%%%%%%%%%%%%

\section{Quantifying and certifying sequential measurement incompatibility}\label{QC}

The measurement incompatibility in quantum theory gives rise to a wide range of intriguing phenomena such as uncertainty relations \cite{Andersson2005, Busch2014}, preparation contextuality \cite{Liang2011, Budroni2022} and state discrimination \cite{Skrzypczyk2019}. Incompatibility is also crucial for comprehending quantum correlations like quantum steering \cite{Uola2020} and nonlocality \cite{Brunner2014, Choudhary2007, Banik2013}. Since incompatible measurements are necessary for demonstrating quantum correlations, sequential violations of the preparation noncontextual bounds of $\mathscr{I}^1$, $\mathscr{I}^2$ and $\mathscr{I}^3$ device-independently certifies that three POVMs for each Bob are incompatible \cite{Tavakoli2020}. It is to be noted here that using the CHSH inequality, two incompatible POVMs have been certified in the sequential scenario in terms of the degree of incompatibility introduced in \cite{Busch2014a}. However, such certification assumed the dimension of the state and observables. Here, by suitably quantifying the degree of incompatibility \cite{Pal2011, Busch2014a, Heinosaari2016}, we first certify the sequential degree of incompatibility for two POVMs in the CHSH scenario. For this purpose, we invoke the recently obtained \cite{Roy2023} DI bounds of CHSH value between Alice-Bob$^1$ and Alice-Bob$^2$. Then, we proceed to the certification of the degree of incompatibility of three POVMs.

\subsection{DI sequential certification of the degree of incompatibility of two POVMs}

Following \cite{Pal2011, Busch2014a}, where qubit observables were assumed, we define the degree of incompatibility between two dichotomic observables in arbitrary dimensions
\begin{equation}
    \mathcal{D}(B_1,B_2) = ||B_1+B_2|| + ||B_1-B_2|| - 2
\end{equation}

All compatible observables obey $\mathcal{D}(B_1,B_2)\leq 0$, whereas $B_0$ and $B_1$ are incompatible, if $0<\mathcal{D}(B_1,B_2)\leq (2\sqrt{2}-2)$.  

Now, considering the following CHSH functional \cite{Clauser1969}
\begin{eqnarray}
    \mathscr{C} = A_1\otimes (B_1+B_2) + A_2 \otimes (B_1-B_2)    
\end{eqnarray}
and employing the SOS method introduced in \cite{Pan2019}, the optimal CHSH value, $\mathcal{C}=\Tr[\mathscr{C}\rho]$, also can be expressed as \cite{Roy2023}
\begin{eqnarray}
    \mathcal{C} \leq ||B_1+B_2|| + ||B_1-B_2|| =  \mathcal{D}(B_1,B_2)+2
\end{eqnarray}
Therefore, $B_1$ and $B_2$ will be incompatible \textit{iff} $ \mathcal{D}(B_1,B_2) \geq \mathcal{C}-2 > 0$.

In the sequential scenario, the degree of incompatibility of Bob$^1$'s observables is given by
\begin{eqnarray}
     \mathcal{D}(B_1^1,B_2^1) \geq \frac{\mathcal{C}^1}{\eta_1} - 2 \label{dicab1}
\end{eqnarray}
Then, the measurements of $B_1^1$ and $B_2^1$ will be incompatible \textit{iff} $\mathcal{C}^1>2$ , which in turn fixes the lower bound of $\eta_1$, i.e., $\frac{1}{\sqrt{2}}<\eta_1 \leq 1$.

The CHSH value between Alice and Bob$^2$ is evaluated as (see Sec.~(3) of \cite{Roy2023})
\begin{eqnarray}
    \mathcal{C}^2 &\leq& \frac{1}{2} \qty(1+\sqrt{1-\eta_1^2}) \ \qty[ ||B_1^2+B_2^2|| + ||B_1^2-B_2^2||] \nonumber \\
    &=& \frac{1}{2} \qty(1+\sqrt{1-\eta_1^2}) \ \qty[ \mathcal{D}(B_1^2,B_2^2)+2]
\end{eqnarray}
Thus, the degree of incompatibility of Bob$^2$'s observables is given by
\begin{eqnarray}
     \mathcal{D}(B_1^2,B_2^2) \geq \frac{2 \mathcal{C}^2}{1+\sqrt{1-\eta_1^2}}-2 \label{dicab2}
\end{eqnarray}
and $B_1^2$ and $B_2^2$ will be incompatible \textit{iff} $\mathcal{C}^2>2$. This in turn fixes the upper bound of $\eta_1$, i.e. $0\leq \eta_1 <\sqrt{2(\sqrt{2}-1)}$.

Now, for obtaining simultaneous CHSH violations between Alice-Bob$^1$ and Alice-Bob$^2$, the degree of incompatibility must be greater than zero, i.e., $\mathcal{D}(B_1^1, B_2^1)>0$ and $\mathcal{D}(B_1^2, B_2^2) > 0$. These conditions then fix both the upper and lower bounds of Bob$^1$'s unsharp parameter $\eta_1$, given by
\begin{equation}
\frac{1}{\sqrt{2}}<\eta_1<\sqrt{2(\sqrt{2}-1)}
\end{equation}

From Eqs.~(\ref{dicab1}) and (\ref{dicab2}), we obtain the trade-off between the degree of incompatibilities of Bob$^1$ and Bob$^2$'s observables. Note that with increasing values of $\eta_1$, while the incompatibility of Bob$^1$'s observables decreases, incompatibility of Bob$^2$'s observable increases up to a certain upper bound of $(\eta_1)_{max}=\sqrt{2 \ \qty(\sqrt{2}-1)}$.

%%%%%%%%%%%%%%%%%%%%%%%%%%%%%%%%%%%%%%%%%%%%%%%%%%%%%%%%%%%%%%%%%%%%%%%%%%%%%%%%%%%%%%%%%%%%%%%%%%%%%%%%%%%%%%%%%%%%%%%%%%%%%%%%%%%%%%%%%%%%%%%%%%%%%%%%%%%%%%%%%%%%%%%%%%%%%%%%%%%%%%%%%%%%%%%%%%%%%%%%%

\subsection{DI sequential certification of the degree of incompatibility of three POVMs}

The degree of incompatibility between any three dichotomic observables is defined as \cite{Pal2011}
\begin{eqnarray}\label{Din4}
    \mathcal{D}(B_1,B_2,B_3) &=& ||B_1+B_2+B_3|| + ||B_1-B_2+B_3|| \nonumber \\
    &+& ||B_1+B_2-B_3|| + ||-B_1+B_2+B_3|| - 4 \nonumber \\
    &&
\end{eqnarray}
Where $B_1$, $B_2$ and $B_3$ are incompatible \textit{iff} $0<\mathcal{D}(B_1,B_2,B_3) \leq 4\qty(\sqrt{3}-1)$. The maximum incompatibility bound is saturated if $\{B_i,B_j\}_{i \neq j} =0$.

However, if we choose $B_1$, $B_2$ and $B_3$ in such a way that they satisfy $B_1+B_2+B_3 =0$ (trine-spin observables), then the degree of incompatibility can be defined as follows
\begin{eqnarray}
    \mathcal{D}_{T}(B_1,B_2,B_3) &=& ||B_1-B_2+B_3|| + ||B_1+B_2-B_3|| \nonumber \\
    &+& ||-B_1+B_2+B_3|| - 4 \label{dito}
\end{eqnarray}

Note that for the case of trine-observables, the maximum value of the right-hand side of Eq.~(\ref{dito}) is $2$ (see Sec.~(\ref{obbf3})). Thus, $B_1$, $B_2$ and $B_3$ are incompatible \textit{iff} $0<\mathcal{D}_T(B_1,B_2,B_3) \leq 2$. The maximum incompatibility bound is saturated if $\{B_i,B_j\}_{i \neq j} =-1$.

In order to witness sequential quantum violations, all Bobs must perform a smeared version of a projective measurement, known as unsharp measurement, characterised by the unbiased POVM. Unbiased POVMs are defined by $\mathcal{B}_y\equiv\qty{\frac{\openone\pm\eta B_y}{2}}$. Consequently, the degree of incompatibility is determined by the unsharpness parameter $\eta$. For three anti-commuting qubit observables $B_y$, $\mathcal{B}_y's$ are jointly measurable or compatible \textit{iff} $\eta \leq \frac{1}{\sqrt{3}}$ \cite{Pal2011}. In addition, it has been demonstrated that the triple-wise joint measurability condition for the smeared version of trine qubit observables is given by $\eta \leq \frac{2}{3}$.

Here, we first certify the sequential measurement incompatibility between a specific class of three observables, known as trine-observables, that satisfy $\sum\limits B_y =0 \ \forall y$. Subsequently, using the sequential scenario considered in \cite{Roy2023}, we will generalise our treatment of the certification of the degree incompatibility for any set of three unbiased POVMs.

\subsubsection{DI sequential certification of the degree of incompatibility of trine observables}

In the sequential scenario, it is evident from Sec.~(\ref{obbf3}) that the degree of incompatibility of Bob$^1$'s observables is given by
\begin{eqnarray}
    \mathcal{D}_T(B^1_1,B^1_2,B^1_3) \geq \frac{\mathscr{I}^1}{\eta_1}-4
\end{eqnarray}
Thus, $B_1^1$, $B_2^1$ and $B_3^1$ are incompatible \textit{iff} $\mathscr{I}^1>4$ which in turn fixes the lower bound of $\eta_1$, i.e., $\frac{2}{3}< \eta_1 \leq 1$.

Reexpressing Eq.~(\ref{nbell}), we arrive at the following  Bell value between Alice-Bob$^2$
\begin{eqnarray}
    \mathscr{I}^2 =\eta_2 \sum\limits_{y=1}^3 \Tr[A_y \otimes \mathfrak{B}_y^2 \ \rho_2]
\end{eqnarray}
where $\mathfrak{B}_1^2=B_1^2+B_2^2-B_3^2$, $\mathfrak{B}_2^1=B_1^2-B_2^2+B_3^2$ and $\mathfrak{B}_3^1=-B_1^2+B_2^2+B_3^2$. By employing the SOS method outlined in Sec.~(\ref{obbf3}), we obtain
\begin{eqnarray}
    \mathscr{I}^2 \leq \eta_2 \sum\limits_{y=1}^3 ||\mathfrak{B}_y^2||_{\rho_2}
\end{eqnarray}
By evaluating $\rho_2$ from Eq.~(\ref{stategen2}), we derive
\begin{eqnarray}
    \mathscr{I}^2 &\leq& \frac{\eta_2}{2} \qty(1+\sqrt{1-\eta_1^2}) \ \sum\limits_{y=1}^3 \ ||\mathfrak{B}^2_y||_{\rho_1} \nonumber \\
    &=& \frac{\eta_2}{2} \qty(1+\sqrt{1-\eta_1^2}) \ \qty[\mathcal{D}_T(B^2_1,B^2_2,B^2_3)+4]
\end{eqnarray}
Thus, the degree of incompatibility of Bob$^2$'s observables is deduced as
\begin{eqnarray}
    \mathcal{D}_T(B^2_1,B^2_2,B^2_3) \geq \frac{2 \mathscr{I}^2}{\eta_2 \qty(1+\sqrt{1-\eta_1^2})}-4 \label{ditb2}
\end{eqnarray}

Hence, it follows from the above Eq.~(\ref{ditb2}) that Bob$^2$'s observables are incompatible \textit{iff} $\mathscr{I}^2>4$. Subsequently, the upper bound of $\eta_1$ is restricted as $0\leq \eta_1 < \frac{2\sqrt{2}}{3}$ and the lower bound of $\eta_2$ is given by $\frac{4}{3\qty(1+\sqrt{1-\eta_1^2})}<\eta_2 \leq 1$, which is in conformity with the bound previously found in Sec.~(\ref{rc}). Therefore, the minimum value of $\eta_2$ is restricted from the lower bound of $\eta_1$, which is $\frac{2}{3}<\eta_2\leq 1$.

Note that both Bobs' observables will be incompatible for the ranges $\frac{2}{3}<\eta_1 \leq \frac{2\sqrt{2}}{3}$ and $(3-\sqrt{5})<\eta_2 \leq 1$. An interesting point to be emphasised here is that although Bob$^2$'s observable are incompatible for $\eta_2 > \frac{2}{3}$, in the sequential nature of the experiment, they become compatible in the range $\frac{2}{3}<\eta_2<(3-\sqrt{5})$. Hence, in the sequential scenario, there exists a trade-off between the degree of incompatibility between Bob$^1$'s and Bob$^2$'s observables.  

Similarly, we analyse the sequential degree of incompatibility of Bob$^3$'s observables. For this, we write the Bell value between Alice-Bob$^3$ in the following way
\begin{eqnarray}
    \mathscr{I}^3 &=& \eta_3 \sum\limits_{y=1}^3 \Tr[A_y \otimes \mathfrak{B}^3_y \ \rho_3] \nonumber \\
    &\leq& \eta_3 \sum\limits_{y=1}^3 \ ||\mathfrak{B}_y^3||_{\rho_3} \nonumber \\
    &\leq & \frac{\eta_3}{4} \qty(1+\sqrt{1-\eta_1^2}) \qty(1+\sqrt{1-\eta_2^2}) \ ||\mathfrak{B}^3_y||_{\rho_1} \nonumber \\
    &=& \frac{\eta_3}{4} \qty(1+\sqrt{1-\eta_1^2}) \qty(1+\sqrt{1-\eta_2^2}) \ \qty[\mathcal{D}_T(B_1^3,B_2^3,B_3^3)+4] \nonumber \\
\end{eqnarray}

Therefore, the degree of incompatibility of Bob$^3$'s observables is given by
\begin{eqnarray}
    \mathcal{D}_T(B_1^3,B_2^3,B_3^3) \geq \frac{4 \mathscr{I}^3}{\eta_3 \qty(1+\sqrt{1-\eta_1^2}) \qty(1+\sqrt{1-\eta_2^2})} -4
\end{eqnarray}
Bob$^3$'s observables are incompatible \textit{iff} $\mathscr{I}^3>4$, which fixes the lower bound of $\eta_3$, upper bounds of $\eta_1$ and $\eta_2$. It is then straightforward to follow that in order to ensure all three sequential Bobs' observable to be incompatible, i.e., $\qty{\mathcal{D}_T (B_1^k, B_2^k, B_3^k)}>0 \ \forall k \in \{1,2,3\}$, the relations between three unsharpness parameters are reproduced as given by Eqs.~(\ref{lambda3b})-(\ref{lambda2b}). 

Notably, three sequential Bobs have the same degree of incompatibility for the following values of unsharpness parameters
\begin{eqnarray}
  &&  \eta_1 = \frac{4 \eta_3\qty(4+\eta_3^2)}{16+12\eta_3^2+\eta_3^4} \ \ ; \ \ \eta_2 = \frac{4\eta_3}{4+\eta_3^2} \nonumber \\
  &&     \frac{1}{2} \qty(3+\sqrt{5}-\sqrt{6\sqrt{5}-2}) \leq \eta_3 \leq 1
\end{eqnarray}
Under such a set of conditions on unsharpness parameters, Bell values between Alice-Bob$^1$, Alice-Bob$^2$ and Alice-Bob$^3$ are identical, given by $\mathscr{I}^k=\frac{24 \eta_3 (4+\eta_3^2)}{16+12\eta_3^2+\eta_3^4}$ $\forall k \in \{1,2,3\}$. This is anticipated because incompatibility implies the Bell violation. If we take $\eta_3=1$, then the values of $\eta_1$ and $\eta_2$ are $\eta_1=\frac{20}{29}$ and $\eta_2 = \frac{4}{5}$, leading to $\mathcal{D}_T (B_1^k,B_2^k,B_3^k) =  \frac{29 \mathscr{I}^k}{20}-4$ $\forall k \in \{1,2,3\}$. Therefore, it is evident that the same degree of incompatibility implies identical Bell values for Alice-Bob$^1$, Alice-Bob$^2$ and Alice-Bob$^3$. If we choose a specific Bell value, such as $\mathscr{I}^k = \frac{120}{29}\approx 4.14$, the degree of incompatibility becomes two, signifying that the optimal Bell value corresponds to the maximum degree of incompatibility.

\begin{figure}[!ht]
\includegraphics[scale=0.4]{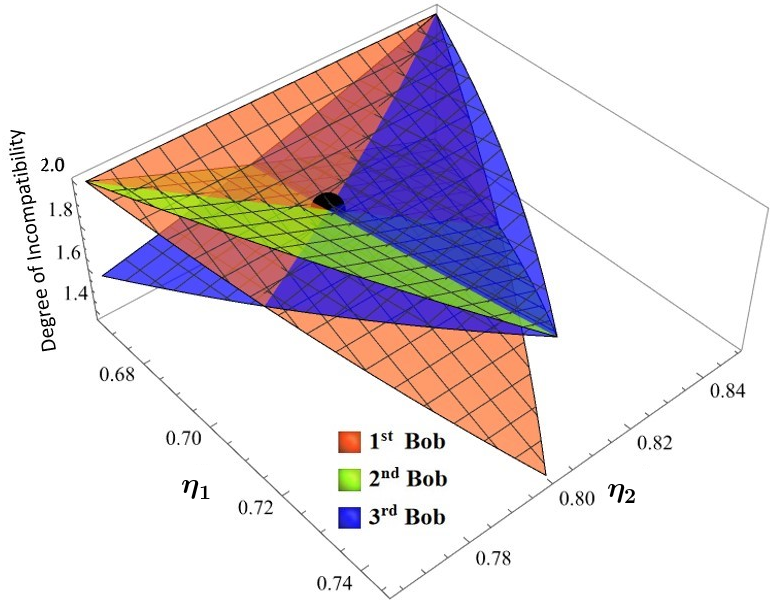}
\caption{The graph illustrates the trade-off between the degree of incompatibilities of three Bobs with respect to $\eta_1$ and $\eta_2$. Here, we take $\eta_3=1$. The yellow, blue and green coloured planes in the 3D plot illustrate variations of the degree of incompatibility of Bob$^1$, Bob$^2$ and Bob$^3$, respectively. The black point on the three-dimensional graph signifies a particular point where the degree of incompatibility of the three Bobs is the same. }
\label{figdoi}
\end{figure}

The variations between the three Bobs degrees of incompatibility are illustrated in Fig.~(\ref{figdoi}).

\section{Summary and discussion}\label{SO}

The present study has explored the self-testing of noisy quantum instruments based on the preparation contextual quantum correlations. For this purpose, we have employed a specific Bell inequality wherein two space-like separated parties, Alice and Bob, perform three measurements each. A notable aspect of such an inequality is that its preparation noncontextual bound of $4$, which is less than the local bound of $5$. Consequently, this offers an advantage in exploiting non-classicality within a quantum correlation. Using an elegant SOS approach, we derived the optimal quantum violation of this inequality to be $6$ devoid of assuming the dimension of the quantum system. Subsequently, based on this optimal bound, the conditions for both party's observables have been established. In particular, the optimal quantum violation self-tests that Alice and Bob's observables must satisfy the following conditions $A_1+A_2+A_3=0$ and $B_1+B_2+B_3=0$ respectively. This in turn provides $\langle\{A_i, A_j\}_{i\neq j}\rangle=\langle\{B_i, B_j\}_{i\neq j}\rangle=-1 \ \forall i,j \in \{1,2,3\}$. This particular class of observables are called trine observables if an analogy is drawn with qubit system. Furthermore, we derived that the shared state must be a maximally entangled state in any dimension. 

We provided the self-testing of unsharp quantum instruments based on the sub-optimal sequential quantum violations of the Bell inequality. We explicitly demonstrated that at most three sequential Bobs can violate the preparation non-contextuality inequality. As mentioned, the standard Bell test is incapable of self-testing of unsharpness parameter as the effect of the unsharp measurements is reflected in the post-measurement states.  Since the sub-optimal quantum violations may originate from many different sources, the self-testing of an unsharp quantum instrument inevitably requires the simultaneous DI certification of the state, measurement, and unsharpness parameter. We showed that the sub-optimal quantum violations of Alice-Bob$^1$, Alice-Bob$^2$, and Alice-Bob$^3$ form an optimal tuple leading to a trade-off relationship among three sequential violations (illustrated in Fig.~\ref{fig1}) and enabling the self-testing of the unsharpness parameters of Bob$^1$ and Bob$^2$. This, in turn, self-tests that the shared state between Alice and Bob$^1$ must be maximally entangled and all observables of Alice and the three sequential Bobs must satisfy the conditions of trine observables. 

Further, we have provided a robust self-testing in practical experimental scenario involving noise and imperfections. Due to the presence of noise, one may obtain less values than the predicted sub-optimal quantum values of Bell functional. We demonstrated that in such a case only specific ranges of unsharp parameters can be self-tested which in turn demonstrates the noise robustness of our protocol. 

Finally, by noting that incompatible measurements are necessary for demonstrating preparation contextual quantum correlations \cite{Tavakoli2020jan, Roy2022}, we have investigated the quantification of sequential Bob's degree of incompatibility. Specifically, we have introduced an expression for the degree of incompatibility of trine class of observables. We then evaluated the lower and upper bounds to demonstrate the extent to which the degree of incompatibility affects the bounds of unsharp parameters. The dependencies of the degree of incompatibility on unsharp parameters are presented in Figs.~\ref{figdoi}.  This dimension-independent analysis is a significant advancement over prior works and opens up new possibilities for comprehending the intricate relationships governing quantum correlations. 

We remark here that a sub-optimal quantum violation of a Bell inequality can occur if the state is not maximally entangled or if both parties' respective measurement operators are not sharp projective measurements. In our present work, we have certified the unsharp parameter by introducing the noise to the measurement. On the other hand, recent studies \cite{Kaniewski2016, Coopmans2019} have demonstrated to what extent a maximally entangled state can be self-tested from the quantum violation of CHSH inequality in the presence of noise in the shared state. While our approach does not consider noise in the state, as sub-optimal sequential violation requires the initial state to be maximally entangled, the works of \cite{Kaniewski2016, Coopmans2019} have not considered the noise in the measurement procedure. Thus, extending our dimension-independent framework, it would be worthwhile to investigate the extent to which one can certify a noisy quantum instrument if a noisy channel affects the initial bipartite state.

%%%%%%%%%%%%%%%%%%%%%%%%%%%%%%%%%%%%%%%%%%%%%%%%%%%%%%%%%%%%%%%%%%%%%%%%%%%%%%%%%%%%%%%%%%%%%%%%%%%%%%%%%%%%%%%%%%%%%%%%%%%%%%%%%%%%%%%%%%%%%%%%%%%%%%%%%%%%%%%%%%%%%%%%%%%%%%%%%%%%%%%%%%%%%%%%%%%%%%%%%%%%%%%%%%%%%%%%%%%%%%%%%%%%%%%%%%%%%%%%%%%%%%%%%%%%%%%%%%%%%%%%%%%%%%%%%%%%%%%%%%%%%%%%%%%%%%%%%%%%%%%%%%%%%%%%%%%%%%%%%%%%%%%%%%%%%%%%%%%%%%%%%%%%%%%%%%%%%%%%%%%%%%%%%%%%%%%%%%%%%%%%%%%%%%%%%%%%%%%%%%%%%%%%%%%%%%%%%%%%%%%%%%%%%%%%%%%%%%%%%%%%%%%%%%%%%%%%%%%%%%%%%%%%

\section*{Acknowledgments}
RP acknowledges the financial support from the Council of Scientific and Industrial Research (CSIR, 09/1001(12429)/2021-EMR-I), Government of India. SS acknowledges the support from the National Natural Science Fund of China (Grant No. G0512250610191) and from the project DST/ICPS/QuST/Theme 1/Q42, Government of India. A.K.P. acknowledges the support
from Research Grant No. SERB/CRG/2021/004258, Government of India.

%%%%%%%%%%%%%%%%%%%%%%%%%%%%%%%%%%%%%%%%%%%%%%%%%%%%%%%%%%%%%%%%%%%%%%%%%%%%%%%%%%%%%%%%%%%%%%%%%%%%%%%%%%%%%%%%%%%%%%%%%%%%%%%%%%%%%%%%%%%%%%%%%%%%%%%%%%%%%%%%%%%%%%%%%%%%%%%%%%%%%%%%%%%%%%%%%%%%%%%%%%%%%%%%%%%%%%%%%%%%%%%%%%%%%%%%%%%%%%%%%%%%%%%%%%%%%%%%%%%%%%%%%%%%%%%%%%%%%%%%%%%%%%%%%%%%%%%%%%%%%%%%%%%%%%%%%%%%%%%%%%%%%%%%%%%%%%%%%%%%%%%%%%%%%%%%%%%%%%%%%%%%%%%%%%%%%%%%%%%%%%%%%%%%%%%%%%%%%%%%%%%%%%%%%%%%%%%%%%%%%%%%%%%%%%%%%%%%%%%%%%%%%%%%%%%%%%%%%%%%%%%%%%%%

\appendix

\begin{widetext}

\section{Derivation of expressions of $C_i\otimes C_i$ and proofs of $[C_i\otimes C_i,C_j\otimes C_j]_{i\neq j}=0$, $\Tr[C_i \ \rho]=1$, $\Tr[(\mathscr{A}_i\otimes B_i) \ \rho]=1$}\label{Co}

$\Tr[(\mathscr{A}_i\otimes B_i) \ \rho]=1$ implies $\rho$ is a pure state. Hence, we rewrite the optimality condition in terms of a pure state $\ket{\psi}$ as
\begin{eqnarray}\label{AiBi}
    \mathscr{A}_i\otimes B_i\ket{\psi}&=&\ket{\psi} \ \ \ \forall i \in \qty{1,2,3}
\end{eqnarray}

Now, for $i=1$, from Eq.~(\ref{AiBi}) we obtain the following set of relations
\begin{eqnarray}
    \mathscr{A}_2\mathscr{A}_1\otimes B_2 B_1\ket{\psi} &=& \mathscr{A}_2\otimes B_2\ket{\psi} \ \ \ \ \ \ \ \  \text{(Multiplying $\mathscr{A}_2\otimes B_2$ from left with Eq.~\ref{AiBi})} \label{p1}\\
    \mathscr{A}_3\mathscr{A}_1\otimes B_3 B_1\ket{\psi}&=&\mathscr{A}_3\otimes B_3\ket{\psi}\label{p2} \ \ \ \ \ \ \ \  \text{(Multiplying $\mathscr{A}_3\otimes B_3$ from left with Eq.~\ref{AiBi}))}\\
    \mathscr{A}_3\mathscr{A}_1\otimes B_2 B_1\ket{\psi}&=&\mathscr{A}_3\otimes B_2\ket{\psi}\label{p3} \ \ \ \ \ \ \ \  \text{(Multiplying $\mathscr{A}_3\otimes B_2$ from left with Eq.~\ref{AiBi})}\\
    \mathscr{A}_2\mathscr{A}_1\otimes B_3 B_1\ket{\psi}&=&\mathscr{A}_2\otimes B_3\ket{\psi}\label{p4} \ \ \ \ \ \ \ \  \text{(Multiplying $\mathscr{A}_2\otimes B_3$ from left with Eq.~\ref{AiBi})}
\end{eqnarray}

First, we consider $C_1\otimes C_1 = \mathscr{A}_1\otimes B_1$, thus implying $\Tr[C_1\otimes C_1 \ \rho]=1$. Next, for $i=2$ and $i=3$, from Eq.~(\ref{AiBi}), we obtain another set of relations 
\begin{eqnarray}
    \mathscr{A}_2\otimes B_3\ket{\psi}&=&\openone\otimes B_3 B_2 \ket{\psi} \ \ \ \ \ \ \ \ \ \ \ \ \ \ \ \text{(Multiplying $\openone\otimes B_3 B_2$ from left with Eq.~\ref{AiBi})} \label{p6} \\
    \mathscr{A}_3\otimes B_2\ket{\psi}&=&\openone\otimes B_2 B_3 \ket{\psi} \ \ \ \ \ \ \ \ \ \ \ \ \ \ \ \text{(Multiplying $\openone\otimes B_2 B_3$ from left with Eq.~\ref{AiBi})} \label{p8}
\end{eqnarray}

Adding $\mathscr{A}_2\otimes B_2\ket{\psi}=\ket{\psi}$ with $\mathscr{A}_3\otimes B_3\ket{\psi}=\ket{\psi}$ and subtracting with Eqs.~(\ref{p6}) and (\ref{p8}), we get
\begin{eqnarray}
    (\mathscr{A}_2\otimes B_2 + \mathscr{A}_3\otimes B_3 - \mathscr{A}_2\otimes B_3 - \mathscr{A}_3\otimes B_2)\ket{\psi} &=& \qty(2 \openone\otimes \openone-\openone\otimes\{B_2,B_3\})\ket{\psi}\nonumber\\
     \frac{1}{3}(\mathscr{A}_2\otimes B_2 + \mathscr{A}_3\otimes B_3 - \mathscr{A}_2\otimes B_3 - \mathscr{A}_3\otimes B_2)\ket{\psi} &=& \ket{\psi} \ \ \ \text{(As $\langle\{B_2,B_3\}\rangle_{\rho} = - 1$)}  \label{p10}\\
     \Tr[\frac{1}{3}(\mathscr{A}_2\otimes B_2 + \mathscr{A}_3\otimes B_3 - \mathscr{A}_2\otimes B_3 - \mathscr{A}_3\otimes B_2) \ \rho] &=& 1 \label{p9}
\end{eqnarray}

Next, taking $C_2\otimes C_2 = \frac{1}{3}(\mathscr{A}_2\otimes B_2 + \mathscr{A}_3\otimes B_3 - \mathscr{A}_2\otimes B_3 - \mathscr{A}_3\otimes B_2)$, Eq.~(\ref{p9}) simplifies to $\Tr[C_2\otimes C_2 \ \rho] = 1$. Finally, adding Eqs.~(\ref{p1}) and  (\ref{p2}) as well as subtracting Eqs.~(\ref{p3})  and (\ref{p4}), we derive
\begin{eqnarray}
    (\mathscr{A}_2\mathscr{A}_1\otimes B_2 B_1 + \mathscr{A}_3\mathscr{A}_1\otimes B_3 B_1 - \mathscr{A}_3\mathscr{A}_1\otimes B_2 B_1 - \mathscr{A}_2\mathscr{A}_1\otimes B_3 B_1)\ket{\psi} &=& (\mathscr{A}_2\otimes B_2 + \mathscr{A}_3\otimes B_3 - \mathscr{A}_2\otimes B_3 - \mathscr{A}_3\otimes B_2)\ket{\psi} \nonumber\\
    (\mathscr{A}_2\mathscr{A}_1\otimes B_2 B_1 + \mathscr{A}_3\mathscr{A}_1\otimes B_3 B_1 - \mathscr{A}_3\mathscr{A}_1\otimes B_2 B_1 - \mathscr{A}_2\mathscr{A}_1\otimes B_3 B_1)\ket{\psi} &=& 3\ket{\psi} \ \ \ \ \ \ \text{(From \ref{p10})} \nonumber\\ 
    %\frac{1}{3}(\mathscr{A}_2\mathscr{A}_1\otimes B_2 B_1 + \mathscr{A}_3\mathscr{A}_1\otimes B_3 B_1 - \mathscr{A}_3\mathscr{A}_1\otimes B_2 B_1 - \mathscr{A}_2\mathscr{A}_1\otimes B_3 B_1)\ket{\psi} &=& \ket{\psi}\nonumber\\ 
    \Tr[\frac{1}{3}(\mathscr{A}_2\mathscr{A}_1\otimes B_2 B_1 + \mathscr{A}_3\mathscr{A}_1\otimes B_3 B_1 - \mathscr{A}_3\mathscr{A}_1\otimes B_2 B_1 - \mathscr{A}_2\mathscr{A}_1\otimes B_3 B_1)] &=& 1\label{p11}
\end{eqnarray}

We choose $C_3\otimes C_3 = \frac{1}{3}(\mathscr{A}_2\mathscr{A}_1\otimes B_2 B_1 + \mathscr{A}_3\mathscr{A}_1\otimes B_3 B_1 - \mathscr{A}_3\mathscr{A}_1\otimes B_2 B_1 - \mathscr{A}_2\mathscr{A}_1\otimes B_3 B_1)]$, consequently $\Tr[C_3\otimes C_3 \ \rho] = 1$. It is straightforward to obtain that $C_3\otimes C_3 = (C_2\otimes C_2)\times (C_1\otimes C_1)$. In the following we show that $[C_i \otimes C_i, C_j \otimes C_j]_{i\neq j}=0 \ \forall i,j \in \{1,2,3\}$.

\begin{eqnarray}\label{c1,c2}
     [C_1\otimes C_1, C_3\otimes C_3] &=&\frac{1}{3} \Big[\mathscr{A}_1\otimes B_1 \ , \ (\mathscr{A}_2\mathscr{A}_1\otimes B_2 B_1 - \mathscr{A}_2\mathscr{A}_1\otimes B_3 B_1 - \mathscr{A}_3\mathscr{A}_1\otimes B_2 B_1+\mathscr{A}_3\mathscr{A}_1\otimes B_3 B_1)\Big] \nonumber \\ 
     &=& \frac{1}{3}\Big\{(\mathscr{A}_1\otimes B_1\nonumber)(\mathscr{A}_2\mathscr{A}_1\otimes B_2 B_1 - \mathscr{A}_2\mathscr{A}_1\otimes B_3 B_1 - \mathscr{A}_3\mathscr{A}_1\otimes B_2 B_1+\mathscr{A}_3\mathscr{A}_1\otimes B_3 B_1) \nonumber\\
     && - (\mathscr{A}_2\mathscr{A}_1\otimes B_2 B_1 - \mathscr{A}_2\mathscr{A}_1\otimes B_3 B_1 - \mathscr{A}_3\mathscr{A}_1\otimes B_2 B_1+\mathscr{A}_3\mathscr{A}_1\otimes B_3 B_1)(\mathscr{A}_1\otimes B_1\nonumber)\Big\} \nonumber \\
     &=&\frac{1}{3} \Big(\mathscr{A}_2 \otimes B_2 + \mathscr{A}_3 \otimes B_3 -\mathscr{A}_3 \otimes B_2 - \mathscr{A}_2 \otimes B_3 - \mathscr{A}_2 \otimes B_2 - \mathscr{A}_3 \otimes B_3 + \mathscr{A}_3 \otimes B_2 + \mathscr{A}_2 \otimes B_3\Big) =0 \nonumber \\
\end{eqnarray}
Here, in order to get the third line from the second line, we invoke the relations $\mathscr{A}_1+\mathscr{A}_2+\mathscr{A}_3=B_1+B_2+B_3=0$ and $\langle\{\mathscr{A}_i,\mathscr{A}_j\}_{i\neq j}\rangle =\langle \{B_i, B_j\}_{i \neq j}\rangle =-1$.
\begin{eqnarray} \label{C1,C3}
    [C_1\otimes C_1,C_2\otimes C_2] &=& \Big[\mathscr{A}_1\otimes B_1,\frac{1}{3}(\mathscr{A}_2\otimes B_2 + \mathscr{A}_3\otimes B_3 - \mathscr{A}_3\otimes B_2 - \mathscr{A}_2\otimes B_3)\Big] \nonumber\\
    &=&\frac{1}{3} \Big\{(\mathscr{A}_1 \mathscr{A}_2 \otimes B_1 B_2 + \mathscr{A}_1 \mathscr{A}_3 \otimes B_1 B_3 - \mathscr{A}_1 \mathscr{A}_3 \otimes B_1 B_2 - \mathscr{A}_1 \mathscr{A}_2 \otimes   B_1 B_3) - (\mathscr{A}_1 \mathscr{A}_2 \otimes B_1 B_2 + \openone\otimes \openone + \openone \otimes B_1 B_2 \nonumber\\
    && +\mathscr{A}_1 \mathscr{A}_2 \otimes \openone + \mathscr{A}_1 \mathscr{A}_3 \otimes B_1 B_3 + \openone\otimes \openone + \mathscr{A}_1 \mathscr{A}_3 \otimes \openone + \openone \otimes B_1 B_3 - \mathscr{A}_1 \mathscr{A}_3 \otimes B_1 B_2 - \openone\otimes \openone - \openone \otimes B_1 B_2 \nonumber\\ 
    && - \mathscr{A}_1 \mathscr{A}_3 \otimes \openone - \mathscr{A}_1 \mathscr{A}_2 \otimes B_1 B_3 - \openone\otimes \openone - \openone \otimes B_1 B_3 - \mathscr{A}_1 \mathscr{A}_2 \otimes \openone)\Big\} = 0
\end{eqnarray}

Therefore, from Eqs.~(\ref{c1,c2}), (\ref{C1,C3}) and given condition $C_3 \otimes C_3= \qty(C_2 \otimes C_2)\cdot\qty(C_1 \otimes C_1)$, it follows that $[C_2\otimes C_2,C_3\otimes C_3] = 0$. Furthermore, $[C_i\otimes C_i,C_j\otimes C_j]_{i\neq j}=0$ implies $\Tr[C_i\otimes C_i \ \rho]=\Tr[C_j\otimes C_j \ \rho] = 1$.

Next, with such construction of $C_i \otimes C_i$ as a function of $\mathscr{A}_y \otimes B_y$, we demonstrate that the optimal conditions of Alice and Bob's observables can be recovered. Let us first recall the expression of the bipartite state $\rho$ from Eq.~(\ref{rhod})
\begin{eqnarray}
    \rho = \frac{1}{d^2} \qty[\openone_d\otimes\openone_d + C_1 \otimes C_1+C_2 \otimes C_2+C_3 \otimes C_3 + \sum_{i=4}^{d^2-1} C_i \otimes C_i]
\end{eqnarray}

Now, in order to proof the relation $\Tr{\mathscr{A}_y\otimes B_y \rho}=1$, we need some additional conditions on $\mathscr{A}_y$ and $B_y$. At optimal condition, these relations are $\{\mathscr{A}_i,\mathscr{A}_j\}_{i\neq j}=\{B_i,B_j\}_{i\neq j}=-\openone_d$ that directly implies $\Tr[B_i.B_j]_{\i\neq j} = \Tr[\mathscr{A}_i.\mathscr{A}_j]_{i\neq j}= -\frac{d}{2}$ and $\Tr[\sum_{i=4}^{d^2-1} \mathscr{A}_jC_i \otimes B_jC_i] =0 \ \ \ \forall j={1,2,3}$.
\begin{eqnarray}
    \Tr[(\mathscr{A}_1\otimes B_1)\rho] &=& \frac{1}{d^2} \Tr[\mathscr{A}_1\otimes B_1 + \openone_d \otimes \openone_d + \frac{1}{3}(\mathscr{A}_2 \otimes B_2 + \mathscr{A}_3 \otimes B_3 -\mathscr{A}_3 \otimes B_2 - \mathscr{A}_2 \otimes B_3+\mathscr{A}_1 \mathscr{A}_2 \otimes B_1 B_2)] \nonumber \\ 
    &&+ \frac{1}{d^2} \Tr[\frac{1}{3}(\mathscr{A}_1 \mathscr{A}_3 \otimes   B_1 B_3 - \mathscr{A}_1 \mathscr{A}_3 \otimes B_1 B_2 - \mathscr{A}_1 \mathscr{A}_2 \otimes   B_1 B_3 )+ \sum_{i=4}^{d^2-1} \mathscr{A}_1C_i \otimes B_1C_i] \nonumber \\
    &=& \frac{1}{d^2} \Tr[\openone_d \otimes \openone_d] + \frac{1}{3 \ d^2}\qty(\frac{d^2}{4}-\frac{d^2}{4}-\frac{d^2}{4}+\frac{d^2}{4})=\frac{1}{d^2} \Tr[\openone_d \otimes \openone_d ] = 1 \label{r1} \\
     \Tr[(\mathscr{A}_2\otimes B_2)\rho] &=& \frac{1}{d^2} \Tr[\mathscr{A}_2\otimes B_2 + \mathscr{A}_2 \mathscr{A}_1\otimes B_2 B_1 + \frac{1}{3}( \openone_d \otimes \openone_d 
 +\mathscr{A}_2 \mathscr{A}_3 \otimes B_2 B_3 - \mathscr{A}_2 \mathscr{A}_3 \otimes \openone_d -\openone_d \otimes B_2 B_3+\mathscr{A}_1  \otimes B_1)] \nonumber \\ 
 &&- \frac{1}{d^2} \Tr[\frac{1}{3}(\mathscr{A}_1 \otimes   B_2 B_3 B_1 - \mathscr{A}_2 \mathscr{A}_3 \mathscr{A}_1 \otimes B_1 +\mathscr{A}_2 \mathscr{A}_3 \mathscr{A}_1 \otimes B_2 B_3 B_1 )+\sum_{i=4}^{d^2-1} \mathscr{A}_2C_i \otimes B_2C_i]\nonumber\\ 
 &=&  \frac{1}{d^2} \qty(\frac{d^2}{4}+\frac{3\ d^2}{4})\Tr[\openone_d \otimes \openone_d ]=\frac{1}{d^2} \Tr[\openone_d \otimes \openone_d ] = 1 \label{r2}\\
 \Tr[(\mathscr{A}_3\otimes B_3)\rho] &=& \frac{1}{d^2} \Tr[\mathscr{A}_3\otimes B_3 + \mathscr{A}_3 \mathscr{A}_1\otimes B_3 B_1 + \frac{1}{3}( \openone_d \otimes \openone_d 
 +\mathscr{A}_3 \mathscr{A}_2 \otimes B_3 B_2 - \mathscr{A}_3 \mathscr{A}_2 \otimes \openone_d -\openone_d \otimes B_3 B_2+\mathscr{A}_1  \otimes B_1)] \nonumber \\ 
 &&- \frac{1}{d^2} \Tr[\frac{1}{3}(\mathscr{A}_1 \otimes   B_3 B_2 B_1 - \mathscr{A}_3 \mathscr{A}_2 \mathscr{A}_1 \otimes B_1 +\mathscr{A}_3 \mathscr{A}_2 \mathscr{A}_1 \otimes B_3 B_2 B_1 )+\sum_{i=4}^{d^2-1} \mathscr{A}_3C_i \otimes B_3C_i] \nonumber \\
 &=& \frac{1}{d^2} \qty(\frac{d^2}{4}+\frac{3\ d^2}{4})\Tr[\openone_d \otimes \openone_d ]=\frac{1}{d^2} \Tr[\openone_d \otimes \openone_d ] = 1 \label{r3}
\end{eqnarray}

%%%%%%%%%%%%%%%%%%%%%%%%%%%%%%%%%%%%%%%%%%%%%%%%%%%%%%%%%%%%%%%%%%%%%%%%%%%%%%%%%%%%%%%%%%%%%%%%%%%%%%%%%%%%%%%%%%%%%%%%%%%%%%%%%%%%%%%%%%%%%%%%%%%%%%%%%%%%%%%%%%%%%%%%%%%%%%%%%%%%%%%%%%%%%%%%%%%%%%%%%%%%%%%%%%%%%%%%%%%%%%%%%%%%%%%%%%%%%%%%%%%%%%%%%%%%%%%%%%%%%%%%%%%%%%%%%%%%%%%%%%%%%%%%%%%%%%%%%%%%%%%%%%%%%%%%%%%%%%%%%%%%%%%%%%%%%%%%%%%%%%%%%%%%%%%%%%%%%%%%%%%%%%%%%%%%%%%%%%%%%%%%%%%%%%%%%%%%%%%%%%%%%%%%%%%%%%%%%%%%%%%%%%%%%%%%%%%%%%%%%%%%%%%%%%%%%%%%%%%%%%%%%%%%

\section{Evaluation of the Bell value between Alice-Bob$^k$} \label{aabk}

The reduced state after $(k-1)^{th}$ Bob's measurement is evaluated as
\begin{eqnarray}
\rho_{k}&=& \frac{1+\xi_{k-1}}{2}  \rho_{k-1} + \frac{1-\xi_{k-1}}{6} \sum_{y=1}^3 (\openone\otimes B_y^{k-1}) \ \rho_{1} \ (\openone\otimes B_y^{k-1}) 
\end{eqnarray}

Now, the Bell value between Alice and Bob$^k$ is given by
\begin{eqnarray} 
\mathscr{I}^k = \Tr[\mathcal{I} \ \rho_k] &=& \eta_k \ \Tr[\bigg\{(A_1+A_2-A_3)\otimes B_1^k+(A_1-A_2+A_3)\otimes B_2^k +(-A_1+A_2+A_3)\otimes B_3^k\bigg\} \ \rho_k] \nonumber \\
&=& \eta_k \sum\limits_{y=1}^{3} \omega_y \Tr[\mathscr{A}_y \otimes B_y^k \ \rho_k] \ \ \ \ 
 \ \ \ \text{[from Eq.~(\ref{omegai})]} \nonumber \\
 &=&\eta_k \ \frac{1+\xi_{k-1}}{2} \sum\limits_{y=1}^{3} \omega_y \Tr[\mathscr{A}_y\otimes B_y^k \ \rho_{k-1}]+\eta_k \ \frac{1-\xi_{k-1}}{6} \sum\limits_{y=1}^{3} \omega_y \Tr[\mathscr{A}_y\otimes \qty(\sum_{y'=1}^3 B_{y'}^{k-1} B_y^k B_{y'}^{k-1}) \ \rho_{k-1}] \nonumber \\
 &=& \eta_k \ \sum\limits_{y=1}^{3} \omega_y \Tr[\mathscr{A}_y\otimes \Tilde{B}_y^k \ \rho_{k-1}] \ \ \ \text{with} \ \ \Tilde{B}_y^k = \frac{1+\xi_{k-1}}{2} \ B_y^k + \frac{1-\xi_{k-1}}{6} \ \sum_{y'=1}^3 B_{y'}^{k-1} B_y^k B_{y'}^{k-1} \ \ \ \forall y \in \{1,2,3\} \label{bellab2a}
\end{eqnarray}

%%%%%%%%%%%%%%%%%%%%%%%%%%%%%%%%%%%%%%%%%%%%%%%%%%%%%%%%

\section{Proof of Eqs.~(\ref{c1}) and (\ref{c2})} \label{aab2c}
 
The condition $\Tilde{B}_1^2+\Tilde{B}_2^2+\Tilde{B}_3^2 = 0 $ implies
\begin{eqnarray}\label{ac1}
    \frac{1+\xi_1}{4}(B_1^2+B_2^2+B_3^3)+\frac{1-\xi_1}{6}\sum_{y,y'=1}^3 B_{y'}^1 B_y^2 B_{y'}^1 = 0 \ \ \ \ \ \ \ \text{[from Eq.~(\ref{bellab2a})]}
\end{eqnarray}

Since $B_y^2$ and $B_i^1B_y^2B_i^1$ are independent of the quantity $\xi_1$, the left-hand side of Equation (\ref{ac1}) can be made zero by ensuring that both the coefficient of $\xi_1$ and $\xi_1^0$ are simultaneously zero. This leads to the following conditions
\begin{eqnarray}\label{ac2}
B_1^2+B_2^2+B_3^3=\frac{2}{3}\sum_{y,y'=1}^3 B_{y'}^1 B_y^2 B_{y'}^1 \ ; \ \ \ \  B_1^2+B_2^2+B_3^3=-\frac{2}{3}\sum_{y,y'=1}^3 B_{y'}^1 B_y^2 B_{y'}^1
\end{eqnarray}

Therefore, Eq.~(\ref{ac2}) implies the following
\begin{eqnarray}
    B_1^2+B_2^2+B_3^3 = 0  \ \ \text{and} \ \ \sum_{y,y'=1}^3 B_{y'}^1 B_y^2 B_{y'}^1 = 0 \label{ac3}
\end{eqnarray}

It is then evident that $ B_1^2+B_2^2+B_3^2 = 0 $ implies $\bigg\langle \{B_i^2,B_j^2\}_{i \neq j}\bigg\rangle_{\rho_1}=-1$.

%%%%%%%%%%%%%%%%%%%%%%%%%%%%%%%%%%%%%%%%%%%%%%%%%%%%%%%%%%%%%%%%%%%%%%%%%%%%%%%%%%%%%%%%%%%%%%%%%%%%%%%%%%%%%%%%%%%%%%%%%%%%%%%%%%%%%%%%%%%%%%%%%%%%%%%%%%%%%%%%%%%%%%%%%%%%%%%%%%%%%%%%%%%%%%%%%%%%%%%%%%%%%%%%%%%%%%%%%%%%%%%%%%%%%%%%%%%%%%%%%%%%%%%%%%%%%%%%%%%%%%%%%%%%%%%%%%%%%%%%%%%%%%%%%%%%%%%%%%%%%%%%%%%%%%%%%%%%%%%%%%%%%%%%%%%%%%%%%%%%%%%%%%%%%%%%%%%%%%%%%%%%%%%%%%%%%%%%%%%%%%%%%%%%%%%%%%%%%%%%%%%%%%%%%%%%%%%%%%%%%%%%%%%%%%%%%%%%%%%%%%%%%%%%%%%%%%%%%%%%%%%%%%%%

\section{Proof of Eq.~(\ref{c3})} \label{aab2c1}

The maximisation condition $\Tilde{\omega}_1^2=\Tilde{\omega}_2^2$ results in $||\Tilde{B}_1^2||_{\rho_1}=||\Tilde{B}_2^2||_{\rho_1}$, implying $\Tr[\Tilde{B^2}_1^{\dagger}\Tilde{B^2}_1\rho_1]=\Tr[\Tilde{B^2}_2^{\dagger}\Tilde{B^2}_2\rho_1]$. Recalling the quantities $\tilde{B}_y^{2}$ from Eq.~(\ref{bellab2a}), we infer the following
\begin{eqnarray}\label{w1w2}
 \sum\limits_{y=1}^2 (-1)^{1+y} \ \qty[ 6(1-\xi_1^2) \ \Bigg\langle \Bigg\{B_y^2,\sum_{{y'}=1}^3 B_{y'}^1 B_y^2 B_{y'}^1\Bigg\}\Bigg\rangle_{\rho_1} + \qty(1-\xi_1)^2 \ \Bigg\langle \Bigg\{\sum_{{y'}=1}^3 B_{y'}^1 B_y^2 B_{y'}^1,\sum_{{y'}=1}^3 B_{y'}^1 B_y^2 B_{y'}^1\Bigg\}\Bigg\rangle_{\rho_1}] = 0 
\end{eqnarray}

Now, comparing the coefficients of $\xi_1$ from both sides of Eq.~(\ref{w1w2}), we get
 \begin{eqnarray}
          \Bigg \langle \Bigg\{\sum_{{y'}=1}^3 B_{y'}^1 B_1^2 B_{y'}^1,\sum_{{y'}=1}^3 B_{y'}^1 B_1^2 B_{y'}^1\Bigg\} \Bigg\rangle_{\rho_1} &=& \Bigg \langle\Bigg\{\sum_{{y'}=1}^3 B_{y'}^1 B_2^2 B_{y'}^1,\sum_{{y'}=1}^3 B_{y'}^1 B_2^2 B_{y'}^1\Bigg\} \Bigg\rangle_{\rho_1} \label{B8}
     \end{eqnarray}

At the maximisation condition of $\mathscr{I}^2$ as stated in Theorem \ref{th1}, we have $\Tilde{\omega}_i^2=\Tilde{\omega}_j^2$ and $\Tilde{B}_1^2+\Tilde{B}_2^2+\Tilde{B}_3^2=0$. This condition satisfies $\Bigg\langle\Bigg\{\frac{\Tilde{B}_i^2}{\Tilde{\omega}_i^2},\frac{\Tilde{B}_j^2}{\Tilde{\omega}_j^2}\Bigg\}_{i\neq j}\Bigg\rangle_{\rho_{1}}=-1$, given that $\Tilde{B}_i^2, \Tilde{B}_j^2$ are not normalised. Hence, we express 
\begin{eqnarray}\label{appc6}
\Bigg\langle\Bigg\{\Tilde{B}_i^2,\Tilde{B}_j^2\Bigg\}_{i\neq j}\Bigg\rangle_{\rho_{1}}= -  ||\Tilde{B}_i^2||^2 = - ||\Tilde{B}_j^2||^2
\end{eqnarray}

Taking $\Bigg\langle\Bigg\{\Tilde{B}_1^2,\Tilde{B}_2^2\Bigg\}\Bigg\rangle_{\rho_{1}}=-||\Tilde{B}_1^2||^2$, we obtain
\begin{eqnarray}\label{appc7}
    &&\frac{1-\xi_1^2}{12}\Bigg\langle\Bigg\{B_1^2,\sum_{y'=1}^3 B_{y'}^1 (B_1^2 + B_2^2) B_{y'}^1\Bigg\}+\Bigg\{B_2^2,\sum_{y'=1}^3 B_{y'}^1 B_1^2 B_{y'}^1\Bigg\}\Bigg\rangle_{\rho_1} + \frac{(1-\xi_1)^2}{36}\Bigg\langle\Bigg\{\sum_{y'=1}^3 B_{y'}^1 B_1^2 B_{y'}^1, \sum_{y'=1}^3 B_{y'}^1 B_2^2 B_{y'}^1\Bigg\}\nonumber\\ 
    && + \Bigg\{\sum_{y'=1}^3 B_{y'}^1 B_1^2 B_{y'}^1, \sum_{y'=1}^3 B_{y'}^1 B_1^2 B_{y'}^1\Bigg\}\Bigg\rangle_{\rho_1} = 0
\end{eqnarray}

In Eq.~(\ref{appc7}), $B_1^2, B_2^2, B_3^2$ are independent of $(\xi_1)^0, (\xi_1)^1, (\xi_1)^2$. Therefore, the coefficients of $(\xi_1)^0, (\xi_1)^1, (\xi_1)^2$ must all be zero. The coefficient $\xi_1=0$ implies
\begin{eqnarray}\label{c8}
    \Bigg\langle\Bigg\{\sum_{y'=1}^3 B_{y'}^1 B_1^2 B_{y'}^1, \sum_{y'=1}^3 B_{y'}^1 B_2^2 B_{y'}^1\Bigg\} + \Bigg\{\sum_{y'=1}^3 B_{y'}^1 B_1^2 B_{y'}^1, \sum_{y'=1}^3 B_{y'}^1 B_1^2 B_{y'}^1\Bigg\}\Bigg\rangle_{\rho_1} = 0
\end{eqnarray}

The expression can be rewritten as
\begin{eqnarray}
    \Bigg\langle\Bigg\{\sum_{y'=1}^3 B_{y'}^1 B_1^2 B_{y'}^1, \sum_{y'=1}^3 B_{y'}^1 (B_1^2 + B_2^2)B_{y'}^1\Bigg\}\Bigg\rangle_{\rho_1} =
   \Bigg\langle\Bigg\{\sum_{y'=1}^3 B_{y'}^1 B_1^2 B_{y'}^1, \sum_{y'=1}^3 B_{y'}^1  B_3^2B_{y'}^1\Bigg\}\Bigg\rangle_{\rho_1} &=& 0
\end{eqnarray}

Likewise, by using $\Bigg\langle\Bigg\{\Tilde{B}_1^2,\Tilde{B}_2^2\Bigg\}\Bigg\rangle_{\rho_{1}}=-||\Tilde{B}_2^2||^2$, we find
\begin{eqnarray}
    \Bigg\langle\Bigg\{\sum_{y'=1}^3 B_{y'}^1 B_2^2 B_{y'}^1, \sum_{y'=1}^3 B_{y'}^1  B_3^2B_{y'}^1\Bigg\}\Bigg\rangle_{\rho_1} = 0
\end{eqnarray}

From other pairs of $\Tilde{B}_1^2, \Tilde{B}_3^2$ and $\Tilde{B}_2^2, \Tilde{B}_3^2$, we obtain
\begin{eqnarray}\label{c11}
    \Bigg\langle\Bigg\{\sum_{y'=1}^3 B_{y'}^1 B_2^2 B_{y'}^1, \sum_{y'=1}^3 B_{y'}^1  B_3^2B_{y'}^1\Bigg\}\Bigg\rangle_{\rho_1} = \Bigg\langle\Bigg\{\sum_{y'=1}^3 B_{y'}^1 B_1^2 B_{y'}^1, \sum_{y'=1}^3 B_{y'}^1  B_3^2B_{y'}^1\Bigg\}\Bigg\rangle_{\rho_1} = \Bigg\langle\Bigg\{\sum_{y'=1}^3 B_{y'}^1 B_1^2 B_{y'}^1, \sum_{y'=1}^3 B_{y'}^1  B_2^2B_{y'}^1\Bigg\}\Bigg\rangle_{\rho_1} = 0
\end{eqnarray}
With the help Eq.~(\ref{c11}) and subsequently squaring Eq.~(\ref{ac2}), we arrive at
\begin{eqnarray}\label{c12}
    \qty(\sum_{y'=1}^3 B_{y'}^1 B_1^2 B_{y'}^1)^2+\qty(\sum_{y'=1}^3 B_{y'}^1 B_2^2 B_{y'}^1)^2+\qty(\sum_{y'=1}^3 B_{y'}^1 B_3^2 B_{y'}^1)^2= 0
\end{eqnarray}

From Eq.~(\ref{B8}), it follows that
\begin{eqnarray}\label{C13}
    \qty(\sum_{y'=1}^3 B_{y'}^1 B_1^2 B_{y'}^1)^2=\qty(\sum_{y'=1}^3 B_{y'}^1 B_2^2 B_{y'}^1)^2=\qty(\sum_{y'=1}^3 B_{y'}^1 B_3^2 B_{y'}^1)^2
\end{eqnarray}

Hence, combining Eqs.~(\ref{c12}) and ~(\ref{C13}), we obtain
\begin{eqnarray}\label{c14}
    \qty(\sum_{y'=1}^3 B_{y'}^1 B_1^2 B_{y'}^1)^2=\qty(\sum_{y'=1}^3 B_{y'}^1 B_2^2 B_{y'}^1)^2=\qty(\sum_{y'=1}^3 B_{y'}^1 B_3^2 B_{y'}^1)^2=0
\end{eqnarray}

Finally, from Eqs.~(\ref{B8}) and Eq.~(\ref{c14}), we deduce
\begin{eqnarray}
    \sum_{y'=1}^3 B_{y'}^1 B_1^2 B_{y'}^1=\sum_{y'=1}^3 B_{y'}^1 B_2^2 B_{y'}^1=\sum_{y'=1}^3 B_{y'}^1 B_3^2 B_{y'}^1 =0 
\end{eqnarray}
%%%%%%%%%%%%%%%%%%%%%%%%%%%%%%%%%%%%%%%%%%%%%%%%%%%%%%%%%%%%%%%%%%%%%%%%%%%%%%%%%%%%%%%%%%%%%%%%%%%%%%%%%%%%%%%%%%%%%%%%%%%%%%%%%%%%%%%%%%%%%%%%%%%%%%%%%%%%%%%%%%%%%%%%%%%%%%%%%%%%%%%%%%%%%%%%%%%%%%%%%%%%%%%%%%%%%%%%%%%%%%%%%%%%%%%%%%%%%%%%%%%%%%%%%%%%%%%%%%%%%%%%%%%%%%%%%%%%%%%%%%%%%%%%%%%%%%%%%%%%%%%%%%%%%%%%%%%%%%%%%%%%%%%%%%%%%%%%%%%%%%%%%%%%%%%%%%%%%%%

\section{Proof of Eq.~(\ref{co3}) and ~(\ref{thm21}) }\label{appI3}

The maximisation of $\mathscr{I}^3$, needs $\sum_{y=1}^3 \Tilde{B}_y^3 = 0$ to be satisfied. This in turn gives the following relationship (refer to Eq.~(\ref{B3f}))
\begin{eqnarray}\label{D1}
    &&\sum_{y=1}^3 \Bigg[\Bigg(\frac{B_y^3}{4}+\sum_{y'=1}^3\frac{B_{y'}^1 B_y^3 B_{y'}^1}{6}+\sum_{y',y''=1}^3\frac{B_{y''}^1 B_{y'}^1 B_y^3 B_{y'}^1 B_{y''}^1}{36}\Bigg)+(\xi_1+\xi_2) \Bigg(\frac{B_y^3}{4}-\sum_{y',y''=1}^3\frac{B_{y''}^1 B_{y'}^1 B_y^3 B_{y'}^1 B_{y''}^1}{36}\Bigg) \nonumber \\
    &&+\xi_1 \xi_2 \Bigg(\frac{B_y^3}{4}+\sum_{y'=1}^3\frac{B_{y'}^1 B_y^3 B_{y'}^1}{6}-\sum_{y',y''=1}^3\frac{B_{y''}^1 B_{y'}^1 B_y^3 B_{y'}^1 B_{y''}^1}{36}\Bigg)\Bigg] = 0
\end{eqnarray}

As $B_y^3$, $B_{y'}^1B_y^2B_{y'}^1$ and $B_{y''}^1 B_{y'}^1 B_y^3 B_{y'}^1 B_{y''}^1$ are independent of the quantities $\xi_1$, $\xi_2$, $\xi_1\xi_2$, the coefficients of $\xi_1$, $\xi_2$, $\xi_1\xi_2$, $(\xi_1\xi_2)^0$ must be zero. This leads to the following relations
\begin{eqnarray}
    \sum_{y=1}^3 \qty(\frac{B_y^3}{4}+\sum_{y'=1}^3\frac{B_{y'}^1 B_y^3 B_{y'}^1}{6}+\sum_{y',y''=1}^3\frac{B_{y''}^1 B_{y'}^1 B_y^3 B_{y'}^1 B_{y''}^1}{36}) = 0 \label{D2} \\
    \sum_{y=1}^3 \qty(\frac{B_y^3}{4}-\sum_{y'=1}^3\frac{B_{y'}^1 B_y^3 B_{y'}^1}{6}+\sum_{y',y''=1}^3\frac{B_{y''}^1 B_{y'}^1 B_y^3 B_{y'}^1 B_{y''}^1}{36}) = 0 \label{D3}\\
    \sum_{y=1}^3 \qty(\frac{B_y^3}{4}-\sum_{y',y''=1}^3\frac{B_{y''}^1 B_{y'}^1 B_y^3 B_{y'}^1 B_{y''}^1}{36}) = 0 \label{D4}
\end{eqnarray}

Subtracting and adding Eqs.~(\ref{D2}) and ~(\ref{D3}), we evaluate the following conditions
\begin{eqnarray}
     \sum_{y,y'=1}^3 B_{y'}^1 B_y^3 B_{y'}^1 = 0\label{D5}\\
     \sum_{y=1}^3 \qty(\frac{B_y^3}{4}+\sum_{y',y''=1}^3\frac{B_{y''}^1 B_{y'}^1 B_y^3 B_{y'}^1 B_{y''}^1}{36}) = 0 \label{D6}
\end{eqnarray}

Adding and subtracting Eqs.~(\ref{D4}) and ~(\ref{D6}) implies
\begin{eqnarray}
    B_1^3+B_2^3+B_3^3 = 0 \label{D7}\\
    \sum_{y,y',y''=1}^3 B_{y''}^1 B_{y'}^1 B_y^3 B_{y'}^1 B_{y''}^1 = 0\label{D8}
\end{eqnarray}

Therefore, it is evident that the condition $ B_1^3+B_2^3+B_3^3 = 0 $ obtained in Eq.~(\ref{D7}) leads to $\bigg\langle \{B_i^3,B_j^3\}_{i \neq j}\bigg\rangle_{\rho_1}=-1$.

%%%%%%%%%%%%%%%%%%%%%%%%%%%%%%%%%%%%%%%%%%%%%%%%%%%%%%%%%%%%%%%%%%%%%%%%%%%%%%%%%%%%%%%%%%%%%%%%%%%%%%%%%%%%%%%%%%%%%%%%%%%%%%%%%%%%%%%%%%%%%%%%%%%%%%%%%%%%%%%%%%%%%%%%%%%%%%%%%%%%%%%%%%%%%%%%%%%%%%%%%%%%%%%%%%%%%%%%%%%%%%%%%%%%%%%%%%%%%%%%%%%%%%%%%%%%%%%%%%%%%%%%%%%%%%%%%%%%%%%%%%%%%%%%%%%%%%%%%%%%%%%%%%%%%%%%%%%%%%%%%%%%%%%%%%%%%%%%%%%%%%%%%%%%%%%%%%%%%%%%%%%%%%%%%%%%%%%%%%%%%%%%%

\section{Proof of Eq.~(\ref{thm24})}\label{appI4}

The maximisation condition $\Tilde{\omega}_1^3=\Tilde{\omega}_2^3$ implies $||\Tilde{B}_1^3||_{\rho_1}=||\Tilde{B}_2^2||_{\rho_1}$, which further leads to $\Tr[\Tilde{B^2}_1^{\dagger}\Tilde{B^2}_1\rho_1]=\Tr[\Tilde{B^2}_2^{\dagger}\Tilde{B^2}_2\rho_1]$. Recalling the quantities $\tilde{B}_y^{3}$ from Eq.~(\ref{B3f}), we arrive at the following expressions
\begin{eqnarray}\label{appE1}
    &&\sum\limits_{y=1}^2 (-1)^{1+y}\Bigg[\frac{(1+\xi_1)(1+\xi_2)(1-\xi_1 \xi_2)}{24}\Bigg\langle\Bigg\{B_y^3,\sum_{y'=1}^3 B_{y'}^1 B_y^3 B_{y'}^1\Bigg\}\Bigg\rangle_{\rho_1}+\frac{(1-\xi_1^2)(1-\xi_2^2)}{144} \Bigg\langle\Bigg\{B_y^3,\sum_{y',y''=1}^3 B_{y''}^1 B_{y'}^1 B_y^3 B_{y'}^1 B_{y''}^1\Bigg\}\Bigg\rangle_{\rho_1} \nonumber \\ 
    &&+ \frac{(1-\xi_1 \xi_2)^2}{36}\Bigg\langle\Bigg\{\sum_{y'=1}^3 B_{y'}^1 B_y^3 B_{y'}^1,\sum_{y'=1}^3 B_{y'}^1 B_y^3 B_{y'}^1\Bigg\}\Bigg\rangle_{\rho_1}  +\frac{(1-\xi_1)(1-\xi_2)(1-\xi_1 \xi_2)}{216} \Bigg\langle\Bigg\{\sum_{y'=1}^3 B_{y'}^1 B_y^3 B_{y'}^1,\sum_{y',y''=1}^3B_{y''}^1 B_{y'}^1 B_y^3 B_{y'}^1 B_{y''}^1\Bigg\}\Bigg\rangle_{\rho_1} \nonumber\\ &&+\frac{(1-\xi_1)^2 (1-\xi_2)^2}{36^2}\Bigg\langle\Bigg\{\sum_{y',y''=1}^3 B_{y''}^1 B_{y'}^1 B_y^3 B_{y'}^1 B_{y''}^1,\sum_{y',y''=1}^3B_{y''}^1 B_{y'}^1 B_y^3 B_{y'}^1 B_{y''}^1\Bigg\}\Bigg\rangle_{\rho_1}\Bigg] = 0
\end{eqnarray} 

Comparing the coefficient of $\xi_1$ and $\xi_1(\xi_2)^2$ from both sides of Eq.~(\ref{appE1}), we get
 \begin{eqnarray}
    &&\sum\limits_{y=1}^2(-1)^{1+y}\Bigg[27\Bigg\langle\Bigg\{B_y^3,\sum_{y'=1}^3 B_{y'}^1 B_y^3 B_{y'}^1\Bigg\}\Bigg\rangle_{\rho_1} - 3\Bigg\langle\Bigg\{\sum_{y'=1}^3 B_{y'}^1 B_y^3 B_{y'}^1,\sum_{y',y''=1}^3B_{y''}^1 B_{y'}^1 B_y^3 B_{y'}^1 B_{y''}^1\Bigg\}\Bigg\rangle_{\rho_1}\Bigg]\nonumber\\ &&= \sum\limits_{y=1}^2 (-1)^{1+y}\Bigg[\Bigg\langle\Bigg\{\sum_{y',y''=1}^3 B_{y''}^1 B_{y'}^1 B_y^3 B_{y'}^1 B_{y''}^1,\sum_{y',y''=1}^3B_{y''}^1 B_{y'}^1 B_y^3 B_{y'}^1 B_{y''}^1\Bigg\}\Bigg\rangle_{\rho_1}\Bigg]
    \label{E1}
\end{eqnarray}
\begin{eqnarray}
    &&\sum\limits_{y=1}^2(-1)^{1+y}\Bigg[-27\Bigg\langle\Bigg\{B_y^3,\sum_{y'=1}^3 B_{y'}^1 B_y^3 B_{y'}^1\Bigg\}\Bigg\rangle_{\rho_1} + 3\Bigg\langle\Bigg\{\sum_{y'=1}^3 B_{y'}^1 B_y^3 B_{y'}^1,\sum_{y',y''=1}^3B_{y''}^1 B_{y'}^1 B_y^3 B_{y'}^1 B_{y''}^1\Bigg\}\Bigg\rangle_{\rho_1}\Bigg]\nonumber\\ &&= \sum\limits_{y=1}^2 (-1)^{1+y}\Bigg[\Bigg\langle\Bigg\{\sum_{y',y''=1}^3 B_{y''}^1 B_{y'}^1 B_y^3 B_{y'}^1 B_{y''}^1,\sum_{y',y''=1}^3B_{y''}^1 B_{y'}^1 B_y^3 B_{y'}^1 B_{y''}^1\Bigg\}\Bigg\rangle_{\rho_1}\Bigg] \label{E2}
\end{eqnarray}

Adding Eqs.~(\ref{E1}) and ~(\ref{E2}), we evaluate
\begin{eqnarray}\label{E3}
    \Bigg\langle\Bigg\{\sum_{y',y''=1}^3 B_{y''}^1 B_{y'}^1 B_1^3 B_{y'}^1 B_{y''}^1,\sum_{y',y''=1}^3B_{y''}^1 B_{y'}^1 B_1^3 B_{y'}^1 B_{y''}^1\Bigg\}\Bigg\rangle_{\rho_1} = \Bigg\langle\Bigg\{\sum_{y',y''=1}^3B_{y''}^1 B_{y'}^1 B_2^3 B_{y'}^1 B_{y''}^1,\sum_{y',y''=1}^3B_{y''}^1 B_{y'}^1 B_2^3 B_{y'}^1 B_{y''}^1\Bigg\}\Bigg\rangle_{\rho_1} 
\end{eqnarray}

Now, comparing the coefficient of $\xi_1\xi_2$ from both sides of Eq.~(\ref{appE1}) and using Eq.~(\ref{E3}), it is straightforward to obtain the following relation
\begin{eqnarray}\label{E4}
    \Bigg\langle\Bigg\{\sum_{y'=1}^3 B_{y'}^1 B_1^3 B_{y'}^1,\sum_{y'=1}^3 B_{y'}^1 B_1^3 B_{y'}^1\Bigg\}\Bigg\rangle_{\rho_1}=\Bigg\langle\Bigg\{\sum_{y'=1}^3 B_{y'}^1 B_2^3 B_{y'}^1,\sum_{y'=1}^3 B_{y'}^1 B_2^3 B_{y'}^1\Bigg\}\Bigg\rangle_{\rho_1}
\end{eqnarray}

At the maximisation condition of $\mathscr{I}^3$ as stated in Theorem \ref{th2}, we have $\Tilde{\omega}_i^3=\Tilde{\omega}_j^3$ and $\Tilde{B}_1^3+\Tilde{B}_2^3+\Tilde{B}_3^3=0$, which satisfies $\Bigg\langle\Bigg\{\frac{\Tilde{B}_i^3}{\Tilde{\omega}_i^3},\frac{\Tilde{B}_j^3}{\Tilde{\omega}_j^3}\Bigg\}_{i\neq j}\Bigg\rangle_{\rho_{1}}=-1$ as $\Tilde{B}_i^3,\Tilde{B}_j^3$ are not normalised. Hence, we deduce
\begin{eqnarray}\label{E5}
\Bigg\langle\Bigg\{\Tilde{B}_i^3,\Tilde{B}_j^3\Bigg\}_{i\neq j}\Bigg\rangle_{\rho_{1}}=-||\Tilde{B}_i^3||^2 =-||\Tilde{B}_j^3||^2 
\end{eqnarray}

Taking  $\Bigg\langle\Bigg\{\Tilde{B}_1^3,\Tilde{B}_2^3\Bigg\}\Bigg\rangle_{\rho_{1}}= \ -||\Tilde{B}_1^3||^2$, we derive the following equation
\begin{eqnarray}\label{E6}
     &&\frac{(1+\xi_1)(1+\xi_2)(1-\xi_1 \xi_2)}{24}\Bigg\langle\Bigg\{B_1^3,\sum_{y'=1}^3 B_{y'}^1 (B_1^3+B_2^3) B_{y'}^1\Bigg\}+\Bigg\{B_2^3,\sum_{y'=1}^3 B_{y'}^1 B_1^3 B_{y'}^1\Bigg\}\Bigg\rangle_{\rho_1}+\frac{(1-\xi_1^2)(1-\xi_2^2)}{144} \nonumber\\
     &&\times \Bigg\langle\Bigg\{B_1^3,\sum_{y',y''=1}^3B_{y''}^1 B_{y'}^1 (B_1^3+B_2^3) B_{y'}^1 B_{y''}^1\Bigg\}+\Bigg\{B_2^3,\sum_{y',y''=1}^3B_{y''}^1 B_{y'}^1 B_1^3 B_{y'}^1 B_{y''}^1\Bigg\}\Bigg\rangle_{\rho_1} +\frac{(1-\xi_1 \xi_2)^2}{36}\Bigg\langle\Bigg\{\sum_{y'=1}^3 B_{y'}^1 B_1^3 B_{y'}^1,\sum_{y'=1}^3 B_{y'}^1 (B_1^3+B_2^3) B_{y'}^1\Bigg\}\Bigg\rangle_{\rho_1} \nonumber \\ 
     &&+ \frac{(1-\xi_1)(1-\xi_2)(1-\xi_1 \xi_2)}{216}  \Bigg\langle\Bigg\{\sum_{y'=1}^3 B_{y'}^1 B_1^3 B_{y'}^1,\sum_{y',y''=1}^3B_{y''}^1 B_{y'}^1 (B_1^3+B_2^3) B_{y'}^1 B_{y''}^1\Bigg\}+\Bigg\{\sum_{y'=1}^3 B_{y'}^1 B_2^3 B_{y'}^1,\sum_{y',y''=1}^3B_{y''}^1 B_{y'}^1 B_1^3 B_{y'}^1 B_{y''}^1\Bigg\} \Bigg\rangle_{\rho_1}\nonumber \\ 
     &&+\frac{(1-\xi_1)^2 (1-\xi_2)^2}{36^2} \Bigg\langle\Bigg\{\sum_{y',y''=1}^3 B_{y''}^1 B_{y'}^1 B_1^3 B_{y'}^1 B_{y''}^1,\sum_{y',y''=1}^3B_{y''}^1 B_{y'}^1 (B_1^3+B_2^3) B_{y'}^1 B_{y''}^1\Bigg\} \Bigg\rangle_{\rho_1} = 0 
\end{eqnarray}

As $\sum_{y=1}^3 B_y^3 =0$, we arrive
\begin{eqnarray}
     &&\frac{(1+\xi_1)(1+\xi_2)(1-\xi_1 \xi_2)}{24}\Bigg\langle\Bigg\{B_2^3,\sum_{y'=1}^3 B_{y'}^1 B_1^3 B_{y'}^1\Bigg\}-\Bigg\{B_1^3,\sum_{y'=1}^3 B_{y'}^1 B_3^3 B_{y'}^1\Bigg\}\Bigg\rangle_{\rho_1}+\frac{(1-\xi_1^2)(1-\xi_2^2)}{144}  \Bigg\langle\Bigg\{B_2^3,\sum_{y',y''=1}^3B_{y''}^1 B_{y'}^1 B_1^3 B_{y'}^1 B_{y''}^1\Bigg\}\nonumber \\ 
     &&-\Bigg\{B_1^3,\sum_{y',y''=1}^3B_{y''}^1 B_{y'}^1 B_3^3 B_{y'}^1 B_{y''}^1\Bigg\}\Bigg\rangle_{\rho_1} -\frac{(1-\xi_1 \xi_2)^2}{36}\Bigg\langle\Bigg\{\sum_{y'=1}^3 B_{y'}^1 B_1^3 B_{y'}^1,\sum_{y'=1}^3 B_{y'}^1 B_3^3 B_{y'}^1\Bigg\}\Bigg\rangle_{\rho_1} + \frac{(1-\xi_1)(1-\xi_2)(1-\xi_1 \xi_2)}{216}  \nonumber \\ 
     &&\times \Bigg\langle\Bigg\{\sum_{y'=1}^3 B_{y'}^1 B_2^3 B_{y'}^1,\sum_{y',y''=1}^3B_{y''}^1 B_{y'}^1 B_1^3 B_{y'}^1 B_{y''}^1\Bigg\}-\Bigg\{\sum_{y'=1}^3 B_{y'}^1 B_1^3 B_{y'}^1,\sum_{y',y''=1}^3B_{y''}^1 B_{y'}^1 B_3^3 B_{y'}^1 B_{y''}^1\Bigg\} \Bigg\rangle_{\rho_1}-\frac{(1-\xi_1)^2 (1-\xi_2)^2}{36^2}\nonumber \\ 
     && \times \Bigg\langle\Bigg\{\sum_{y',y''=1}^3 B_{y''}^1 B_{y'}^1 B_1^3 B_{y'}^1 B_{y''}^1,\sum_{y',y''=1}^3B_{y''}^1 B_{y'}^1 B_3^3 B_{y'}^1 B_{y''}^1\Bigg\} \Bigg\rangle_{\rho_1} = 0 
\end{eqnarray}

In Eq.~(\ref{E6}), as $B_1^3, B_2^3,B_3^3$ are independent of $\xi_2, \xi_1, \xi_1(\xi_2)^2, \xi_1\xi_2$, the coefficients of $\xi_2, \xi_1, \xi_1(\xi_2)^2, \xi_1\xi_2$ must all be equal to zero. From Eq.~(\ref{E6}), coefficient of $\xi_1=0$ implies
\begin{eqnarray}\label{E7}
    &&27 \Bigg\langle\Bigg\{B_1^3,\sum_{y'=1}^3 B_{y'}^1 B_3^3 B_{y'}^1\Bigg\}-\Bigg\{B_2^3,\sum_{y'=1}^3 B_{y'}^1 B_1^3 B_{y'}^1\Bigg\}\Bigg\rangle_{\rho_1} - \ 3\Bigg\langle\Bigg\{\sum_{y'=1}^3 B_{y'}^1 B_1^3 B_{y'}^1,\sum_{y',y''=1}^3B_{y''}^1 B_{y'}^1 B_3^3 B_{y'}^1 B_{y''}^1\Bigg\} \nonumber\\&& - \Bigg\{\sum_{y'=1}^3 B_{y'}^1 B_2^3 B_{y'}^1,\sum_{y',y''=1}^3B_{y''}^1 B_{y'}^1 B_1^3 B_{y'}^1 B_{y''}^1\Bigg\} \Bigg\rangle_{\rho_1} = \ \Bigg\langle\Bigg\{\sum_{y',y''=1}^3 B_{y''}^1 B_{y'}^1 B_1^3 B_{y'}^1 B_{y''}^1,\sum_{y',y''=1}^3B_{y''}^1 B_{y'}^1 B_3^3 B_{y'}^1 B_{y''}^1\Bigg\}\Bigg\rangle_{\rho_1}
\end{eqnarray}

From Eq.~(\ref{E6}), coefficient of $\xi_1 (\xi_2)^2=0$ implies
\begin{eqnarray}\label{E8}
    &&-27 \Bigg\langle\Bigg\{B_1^3,\sum_{y'=1}^3 B_{y'}^1 B_3^3 B_{y'}^1\Bigg\}-\Bigg\{B_2^3,\sum_{y'=1}^3 B_{y'}^1 B_1^3 B_{y'}^1\Bigg\}\Bigg\rangle_{\rho_1} + \ 3\Bigg\langle\Bigg\{\sum_{y'=1}^3 B_{y'}^1 B_1^3 B_{y'}^1,\sum_{y',y''=1}^3B_{y''}^1 B_{y'}^1 B_3^3 B_{y'}^1 B_{y''}^1\Bigg\} \nonumber\\
    && - \Bigg\{\sum_{y'=1}^3 B_{y'}^1 B_2^3 B_{y'}^1,\sum_{y',y''=1}^3B_{y''}^1 B_{y'}^1 B_1^3 B_{y'}^1 B_{y''}^1\Bigg\} \Bigg\rangle_{\rho_1} = \ \Bigg\langle\Bigg\{\sum_{y',y''=1}^3 B_{y''}^1 B_{y'}^1 B_1^3 B_{y'}^1 B_{y''}^1,\sum_{y',y''=1}^3B_{y''}^1 B_{y'}^1 B_3^3 B_{y'}^1 B_{y''}^1\Bigg\}\Bigg\rangle_{\rho_1}
\end{eqnarray}

Adding Eqs.~(\ref{E7}) and ~(\ref{E8}), we derive
\begin{eqnarray}\label{E9}
    &&\Bigg\langle \Bigg\{\sum_{y',y''=1}^3 B_{y''}^1 B_{y'}^1 B_1^3 B_{y'}^1 B_{y''}^1,\sum_{y',y''=1}^3 B_{y''}^1 B_{y'}^1 B_3^3 B_{y'}^1 B_{y''}^1\Bigg\} \Bigg \rangle_{\rho_1} = 0 
\end{eqnarray}

Now, following the similar procedure for obtaining Eq.~(\ref{E9}), and taking into account the other pairs $B_1^3, B_2^3$ and $B_2^3, B_2^3$, we arrive at the following condition
\begin{eqnarray}\label{E10}
    && \Bigg\langle \Bigg\{\sum_{y',y''=1}^3 B_{y''}^1 B_{y'}^1 B_m^3 B_{y'}^1 B_{y''}^1,\sum_{y',y''=1}^3 B_{y''}^1 B_{y'}^1 B_n^3 B_{y'}^1 B_{y''}^1\Bigg\} \Bigg \rangle_{\rho_1} =0 \ \ \ \ \ \text{$m\neq n \ , \{m,n\}\in \{1,2,3\}$}
\end{eqnarray}

Comparing the coefficients of $\xi_1 \xi_2$ from Eq.~(\ref{E6}) and putting it into Eq.~(\ref{E9}), we obtain
\begin{eqnarray}\label{E11}
    \Bigg\langle\Bigg\{\sum_{y'=1}^3 B_{y'}^1 B_1^3 B_{y'}^1,\sum_{y'=1}^3 B_{y'}^1 (B_1^3+B_2^3) B_{y'}^1\Bigg\}\Bigg\rangle_{\rho_1} = 0 \ \ ; \ \ \Bigg\langle\Bigg\{\sum_{y'=1}^3 B_{y'}^1 B_1^3 B_{y'}^1,\sum_{y'=1}^3 B_{y'}^1 B_3^3 B_{y'}^1\Bigg\}\Bigg\rangle_{\rho_1} = 0 
\end{eqnarray}

Likewise Eq.~(\ref{E11}), considering other pairs of $B_m^3$, $B_n^3$, we derive
\begin{eqnarray}\label{E12}
    \Bigg\langle\Bigg\{\sum_{y'=1}^3 B_{y'}^1 B_m^3 B_{y'}^1,\sum_{y'=1}^3 B_{y'}^1 B_n^3 B_{y'}^1\Bigg\}\Bigg\rangle_{\rho_1} = 0 \ \ \ \ \ \text{$m\neq n \ , \{m,n\}\in \{1,2,3\}$}
\end{eqnarray}

using Eq.~(\ref{E12}) and squaring Eq.~(\ref{D5}), we deduce
\begin{eqnarray}\label{E13}
    \Bigg\langle\Bigg(\sum_{y'=1}^3 B_{y'}^1 B_1^3 B_{y'}^1\Bigg)^2+\Bigg(\sum_{y'=1}^3 B_{y'}^1 B_2^3 B_{y'}^1\Bigg)^2+\Bigg(\sum_{y'=1}^3 B_{y'}^1 B_3^3 B_{y'}^1\Bigg)^2\Bigg\rangle_{\rho_1} = 0
\end{eqnarray}

Now from Eqs.~(\ref{E4}) and (\ref{E13}), it is evident that
\begin{eqnarray}\label{E14}
    \Bigg\langle\Bigg(\sum_{y'=1}^3 B_{y'}^1 B_1^3 B_{y'}^1\bigg)^2\Bigg\rangle_{\rho_1}=\Bigg\langle\Bigg(\sum_{y'=1}^3 B_{y'}^1 B_2^3 B_{y'}^1\Bigg)^2\Bigg\rangle_{\rho_1}=\Bigg\langle\Bigg(\sum_{{y'}=1}^3 B_{y'}^1 B_3^3 B_{y'}^1\Bigg)^2\Bigg\rangle_{\rho_1}
\end{eqnarray}

Hence, from Eqs.~(\ref{E13}) and (\ref{E14}), it is straight forward to obtain that
\begin{eqnarray}\label{15}
 \Bigg\langle\sum_{y'=1}^3 B_{y'}^1 B_y^2 B_{y'}^1\Bigg\rangle_{\rho_1} &=& 0  \ \ \forall y \in \{1,2,3\} 
 \end{eqnarray}
 
With the help of Eq.~(\ref{E10}) and squaring of Eq.~(\ref{D8}), we evaluate
 \begin{eqnarray}\label{E16}
     \Bigg\langle\Bigg(\sum_{y',y''=1}^3B_{y''}^1 B_{y'}^1 B_1^3 B_{y'}^1 B_{y''}^1\Bigg)^2+\Bigg(\sum_{y',y''=1}^3B_{y''}^1 B_{y'}^1 B_2^3 B_{y'}^1 B_{y''}^1\Bigg)^2+\Bigg(\sum_{y',y''=1}^3B_{y''}^1 B_{y'}^1 B_3^3 B_{y'}^1 B_{y''}^1\Bigg)^2\Bigg\rangle_{\rho_1} = 0
 \end{eqnarray}

Next, Eqs.~(\ref{E3}) and (\ref{E16}) leads to the following relation
 \begin{eqnarray}\label{E17}
     \Bigg\langle\Bigg(\sum_{y',y''=1}^3B_{y''}^1 B_{y'}^1 B_1^3 B_{y'}^1 B_{y''}^1\Bigg)^2\Bigg\rangle_{\rho_1}=\Bigg\langle\Bigg(\sum_{y',y''=1}^3B_{y''}^1 B_{y'}^1 B_2^3 B_{y'}^1 B_{y''}^1\Bigg)^2\Bigg\rangle_{\rho_1}=\Bigg\langle\Bigg(\sum_{y',y''=1}^3B_{y''}^1 B_{y'}^1 B_3^3 B_{y'}^1 B_{y''}^1\Bigg)^2)^2\Bigg\rangle_{\rho_1}
 \end{eqnarray}

Finally, from Eqs.~(\ref{E16}) and (\ref{E17}), we show that
 \begin{eqnarray}
     \Bigg\langle\Bigg(\sum_{y',y''=1}^3B_{y''}^1 B_{y'}^1 B_y^3 B_{y'}^1 B_{y''}^1\Bigg)\Bigg\rangle_{\rho_1} = 0  \ \ \forall y \in \{1,2,3\}
 \end{eqnarray}
\end{widetext}

%%%%%%%%%%%%%%%%%%%%%%%%%%%%%%%%%%%%%%%%%%%%%%%%%%%%%%%%%%%%%%%%%%%%%%%%%%%%%%%%%%%%%%%%%%%%%%%%%%%%%%%%%%%%%%%%%%%%%%%%%%%%%%%%%%%%%%%%%%%%%%%%%%%%%%%%%%%%%%%%%%%%%%%%%%%%%%%%%%%%%%%%%%%%%%%%%%%%%%%%%%%%%%%%%%%%%%%%%%%%%%%%%%%%%%%%%%%%%%%%%%%%%%%%%%%%%%%%%%%%%%%%%%%%%%%%%%%%%%%%%%%%%%%%%%%%%%%%%%%%%%%%%%%%%%%%%%%%%%%%%%%%%%%%%%%%%%%%%%%%%%%%%%%%%%%%%%%%%%%%%%%%%%%%%%%%%%%%%%%%%%%%%%%%%%%%%%%%%%%%%%%%%%%%%%%%%%%%%%%%%%%%%%%%%%%%%%%%%%%%%%%%%%%%%%%%%%%%%%%%%%%%%%%%

 %\section*{References}
 %\bibliographystyle{JHEP}
 \bibliography{references}

\end{document}